\documentstyle[aps,prd,epsfig]{revtex}

\begin{document}

\def\Journal#1&#2&#3(#4){#1{\bf #2} (#4) #3}

% Some useful journal names
\def\AP{{\em Ann. Phys. }}
\def\NIM{{\em Nucl. Inst. and Meth. }}
\def\JPG{{\em J. Phys. }{\bf G}}
\def\NIMA{{\em Nucl. Inst. and Meth. }{\bf A}}
\def\NCA{{\em Nuovo. Cim. }{\bf A}}
\def\JETFL{{\em JETF Lett. }}
\def\NPA{{\em Nucl. Phys. }{\bf A}}
\def\NPB{{\em Nucl. Phys. }{\bf B}}
\def\PLB{{\em Phys. Lett. }{\bf B}}
\def\PL{{\em Phys. Lett. }}
\def\PRL{{\em Phys. Rev. Lett. }}
\def\TMP{{\em Theor. Math. Phys. }}
\def\PR{{\em Phys. Rep. }}	
\def\PRD{{\em Phys. Rev. }{\bf D}}
\def\PREV{{\em Phys. Rev. }}
\def\PRC{{\em Phys. Rep. }{\bf C}}
\def\ZPA{{\em Z. Phys. }{\bf A}}
\def\ZPC{{\em Z. Phys. }{\bf C}}
\def\IJMF{{\em Int. J. Mod. Phys. }}

\def\earc#1&#2(#3){#1{\bf #2} (#3).}

\def\newblock{~}

\def\hepph  {{\rm hep-ph }}

\def\etal   {{\it et al.}}
\def\etaln  {{\it et al}}

\def\gevt   {{\rm GeV^2}}

\def\piz    {\pi^0}
\def\etaz   {\eta}
\def\etap   {\eta^{\prime}}
\def\ra     {\rightarrow}
\def\gam    {\gamma}
\def\pip    {\pi^+}
\def\pim    {\pi^-}
\def\rhoz   {\rho^0}
\def\sig    {\sigma}
\def\qsq    {Q^2}

\def\qsqp   {{Q^\prime}^2}

\def\gaga   {\gamma\gamma}
\def\gsg    {\gamma^*\gamma}
\def\mgsg   {\gamma^*\gamma\rightarrow{\cal R}}
\def\mgsgs  {\gamma^*\gamma^*\rightarrow{\cal R}}
\def\gsgs   {\gamma^*\gamma^*}
\def\qq     {q\tilde{Q}}

\def\frg    {{\cal F}_{\gamma^*\gamma{\cal R}}}
\def\frgs   {{\cal F}_{\gamma^*\gamma^*{\cal R}}}
\def\fpizg  {{\cal F}_{\gamma^*\gamma\pi^0}}
\def\fpizgs {{\cal F}_{\gamma^*\gamma^*\pi^0}}
\def\fetazg {{\cal F}_{\gamma^*\gamma\eta}}
\def\fetapg {{\cal F}_{\gamma^*\gamma\eta^{\prime}}}

\title{\Large \bf  
\rightline{\normalsize\rm  CLNS 97/1477}
\rightline{\normalsize\rm  CLEO 97-7}
\rightline{}
\rightline{}
\centerline{\Large\bf Measurements of the Meson-Photon Transition}
\centerline{\Large\bf Form Factors of Light Pseudoscalar Mesons} 
\centerline{\Large\bf at Large Momentum Transfer}
}

\vspace{0.2cm}
\author{CLEO Collaboration}
\vspace{0.2cm}
%\date{April 26, 1997}
\date{\today}

\maketitle
\tighten

\bigskip

\begin{abstract}
Using the CLEO~II detector, we 
have measured the differential cross sections 
for exclusive two-photon production 
of light pseudoscalar mesons $\pi^0$, $\eta$, and $\eta^{\prime}$. 
From our measurements we have obtained the 
form factors associated with the 
electromagnetic transitions 
$\gamma^*\gamma$ $\rightarrow$ meson. 
We have measured these form factors 
in the momentum transfer ranges 
from 1.5 to 
$9$, $20$, and $30$ GeV$^2$ 
for $\pi^0$, $\eta$, and $\eta^{\prime}$, 
respectively, 
and have made comparisons to various theoretical predictions. 
\vspace{3mm}

\noindent{12.38.Qk,13.40.Gp,13.65.+i}
\end{abstract} 

% 12.38.Qk Experimental tests (of QCD)
% 13.40.Gp Electromagnetic form factors
% 13.65.+i Hadron production by electron-positron collisions
%%\pacs{12.38.Qk,13.40.Gp,13.65.+i}

% Insert author and address list here

\newpage 

\setcounter{footnote}{0}
\renewcommand{\thefootnote}{\alph{footnote}}

\begin{center}
J.~Gronberg,$^{1}$ T.~S.~Hill,$^{1}$ R.~Kutschke,$^{1}$
D.~J.~Lange,$^{1}$ S.~Menary,$^{1}$ R.~J.~Morrison,$^{1}$
H.~N.~Nelson,$^{1}$ T.~K.~Nelson,$^{1}$ C.~Qiao,$^{1}$
J.~D.~Richman,$^{1}$
% ...
\end{center}
 
\small
\begin{center}
$^{1}${University of California, Santa Barbara, California 93106}\\
$^{2}${University of Colorado, Boulder, Colorado 80309-0390}\\
% ...
$^{12}${University of Minnesota, Minneapolis, Minnesota 55455}\\
$^{13}${State University of New York at Albany, Albany, New York 12222}\\
% ...
\end{center}

% ----------------Start Introduction here ------------------------------------

\setcounter{footnote}{0}
\renewcommand{\thefootnote}{\arabic{footnote}}

\newpage 

\section{Introduction}

Production of even $C$-parity hadronic matter in $e^+e^-$ scattering provides 
a unique opportunity to study the properties of strong interactions. 
To leading order in quantum electrodynamics (QED) these processes 
are described as the interaction between two photons 
emitted by the scattered electrons\footnote
{Unless otherwise specified, 
we use the term ``electron'' for either an electron or a positron.}. 
Although in $e^+e^-$ scattering the probe and the target are both 
represented by photons that are carriers of the electromagnetic force, 
these space-like photons can produce a pair of quarks 
that interact strongly and are observed in the form of hadrons. 
Therefore, by measuring the four-momenta of the 
scattered electrons we can study the 
dynamics of strong interactions. 
The quantities of interest in these studies 
are the form factors associated with the 
transitions between the photons and the hadrons. 

This Article describes the measurements~\cite{SAVINOV:thesis} of 
the differential cross sections for the production of a single pseudoscalar 
meson in $e^+e^-$ scattering: 
\begin{equation}
%\[
e^+e^- \ra e^+e^-{\cal R}, 
%\]
\label{EQ:1}
\end{equation} 
where ${\cal R}$ is a $\piz$, $\etaz$ or $\etap$. 
We measure these cross sections in a ``single-tagged'' experimental mode 
where one of the scattered electrons is detected (``tagged''), 
while the other electron is scattered at a very small angle and 
therefore remains undetected (``{\em un}tagged''). 
The mesons produced in $e^+e^-$ scattering are observed 
through their decays to various fully reconstructed final states. 
The tagged electron 
emits a highly off-shell photon ($\gamma^*$), whereas the untagged 
electron emits 
a nearly on-shell photon ($\gamma$). We measure the dependence of the meson 
production rate on the squared momentum transfer $Q^2$ carried by 
the highly off-shell photon. 
This momentum transfer is determined by energy-momentum conservation as applied to 
the tag: 
\begin{equation}
%\[
Q^2 \equiv 
-( p_b - p_t)^2 = 
2 E_b E_t (1-\cos\theta_t),
%\] 
\label{EQ:12}
\end{equation} 
where $p_b$ and $p_t$ are the four-momenta 
of the incident 
beam-energy electron and the tag, 
$E_b$ and $E_t$ are corresponding energies, 
and $\theta_t$ is the scattering angle\footnote
{
The electron mass is neglected in Eqn.~\ref{EQ:12}.}. 
From the measurements of the differential rates 
\begin{equation}
%\[
\frac{d\sigma(e^+e^- \ra e^+e^-{\cal R})}{d\qsq} 
%\]
\label{EQ:13}
\end{equation} 
we obtain the transition form factors $\frg$ 
that describe the effect of the strong interaction in the 
$\mgsg$ transition amplitudes. 

To relate the differential cross sections to 
the transition form factors we employ the theoretical 
framework developed by V.M. Budnev \etal ~\cite{BGMS:75} (BGMS formalism). 
In BGMS the process $e^+e^- \ra e^+e^-{\cal R}$ is divided 
into two parts: $e^+e^- \ra e^+e^-\gamma^*\gamma$ and 
$\gamma^*\gamma \ra {\cal R}$. 
The first part is completely calculable in QED and the second part is 
defined in terms of the transition form factors $\frg(\qsq)$. 
In the case of pseudoscalar mesons there is only one form factor. 
At zero momentum transfer this form factor is expressed as: 
\begin{equation}
%\[
|\frg(0)|^2 =\,
\frac{1}{(4 \pi \alpha)^2} \,
\frac{64 \pi \Gamma(\cal R \ra \gaga)}{M_{\cal R}^3} \,
, 
%\]
\label{EQ:37}
\end{equation} 		
where $\alpha$ is the QED coupling constant, 
$M_{\cal R}$ is the mass 
and 
$\Gamma(\cal R \ra \gamma\gamma)$ 
is 
the two-photon partial width of the meson ${\cal R}$. 
The transition form factors cannot be calculated directly from 
Quantum Chromodynamics (QCD). 
However, they have been estimated using 
perturbative QCD (PQCD), 
a sum-rules approach, 
and other theoretical methods. 

One of the important concepts of PQCD-based methods  
is a factorization procedure that 
separates perturbative short-distance effects 
from non-perturbative long-distance ones. 
While the former are understood well and 
can be calculated using PQCD, 
the latter are known only asymptotically, in the limit $\qsq\ra\infty$. 
In PQCD-based calculations the transition form factor $\frg$ 
is expressed as a convolution 
of a perturbative Hard Scattering Amplitude (HSA)~\cite{BL:80} 
and the soft non-perturbative wave function of the meson. 

Brodsky and Lepage employed PQCD to find 
the asymptotic behavior of the $\mgsg$ transition 
form factors in the limit $\qsq\ra\infty$~\cite{BL:81}: 
\begin{equation} 
%\[ 
\lim_{\qsq \ra \infty} \qsq \frg(\qsq) = 2 f_{\cal R}, 
%\] 
\label{EQ:41} 
\end{equation} 
where $f_{\cal R}$ is the meson decay constant. 
In addition, it has been predicted  
that in this limit any mesonic wave function 
evolves to the asymptotic wave function of unique shape~\cite{BL:80,ER:79,CZ:84}. 

While PQCD predicts the form factors of the $\mgsg$ transitions at large momentum transfer, 
the behavior of these form factors in the limit $\qsq\ra0$ 
can be determined from the axial anomaly~\cite{ANOMALY:axial1,ANOMALY:axial2} 
in the chiral limit of QCD. 
For $\piz$ and $\etaz$ the axial anomaly yields~\cite{BL:81}: 
\begin{equation}
%\[
\lim_{\qsq \ra 0} \frg(\qsq) = \frac{1}{4 \pi^2 f_{\cal R}}, 
%\]
\label{EQ:319}
\end{equation} 
to leading order in $m_u^2/M_{\cal R}^2$ and $m_d^2/M_{\cal R}^2$ where 
$m_u$ and $m_d$ are the masses of the $u$ and $d$ quarks. 
This prediction does not hold with the same precision 
for $\etap$ due to the larger value of the $s$-quark mass. 
In addition, even if the $s$-quark mass were small, 
this prediction might be broken for $\etap$ 
because this particle is an unlikely candidate for the Goldstone boson~\cite{RYDER:U1,COLEMAN:U1}. 

To describe the soft non-perturbative region of $\qsq$ 
a simple interpolation between $\qsq \ra 0$ and $\qsq \ra \infty$ limits has been 
proposed~\cite{BL:81}: 
\begin{equation}
%\[
\frg(\qsq) \sim \frac{1}{4 \pi^2 f_{\cal R}} \frac{1}{1+(\qsq/8 \pi^2 f_{\cal R}^2)}. 
%\]
\label{EQ:45}
\end{equation} 

To quantify the long-distance effects in the soft non-perturbative region, 
Chernyak and Zhitnitsky 
employed the sum-rules method~\cite{AV:sum_rules} to derive 
the wave function of the pion at experimentally accessible momentum 
transfers (the CZ wave function)~\cite{CZ:84}. 
They demonstrated that the proposed wave function 
successfully describes experimental data on the $\chi_c$ 
decay into two pions and the electromagnetic form factor of the charged pion. 
However, because the theoretical predictions for these processes depend on 
the strong interaction coupling constant $\alpha_s$, 
this introduced a large uncertainty in the determination of the CZ wave function. 

Since the asymptotic and CZ wave functions were proposed, they 
have often been used to describe the non-perturbative 
parts of transition amplitudes in various PQCD calculations. 
Jakob, Kroll, and Raulfs employed these 
wave functions and PQCD to calculate 
$\fpizg$~\cite{KROLL:WUB9417,KROLL:96}. 
These authors have also taken into account small QCD radiative corrections, 
incorporated into the PQCD technique by Lee and Sterman~\cite{LS:92}.  
Kroll has concluded that the CZ wave function 
disagrees with our preliminary results~\cite{CLEO:PH95:FF}. 
On the contrary, a competing perturbative analysis of 
Cao, Huang, and Ma~\cite{GUANG:96} yielded that 
either the asymptotic or the CZ wave function is sufficient 
to describe the data. 
These authors took into account quark transverse 
momentum corrections and neglected the QCD radiative corrections, 
estimating the latter as small. 

While PQCD-based methods are often employed to predict rates 
for exclusive processes\footnote{ 
For example, these methods have been utilized to calculate 
the nucleon form factors~\cite{LI:93,BJK_neutron:95} and 
the $\bar{B^0} \ra \pip\pim$ branching fraction~\cite{DJK:95}; 
see also~\cite{BJK_proton:95,CHIBISOV:95}.}, 
the applicability of these methods at experimentally accessible momentum transfers remains 
one of the outstanding problems of the theory of strong interactions. 
Extensive discussion of the validity of the PQCD approach can be found in the 
literature~\cite{ILS:84,BS:84,ILS:89,KROLL:93,BRAUN:94,AVR:CEBAF94,AZ:9605226}. 

To avoid ambiguities of the PQCD-inspired calculations 
at $\qsq$ of the order of several $\gevt$, 
Radyushkin \etal ~developed an approach~\cite{ER:80,RR:91,MR:9702443} 
based on the sum-rules method~\cite{AV:sum_rules} that they employed to 
predict the $\gamma^*\gamma \ra \piz$ transition form factor~\cite{RR:96}. 
This prediction depends on the model of the hadronic spectrum 
chosen to describe an almost real photon emitted by the untagged electron. 
It also depends on the values of vacuum condensates which represent non-perturbative 
matrix elements. 
The theoretical result of Radyushkin \etal ~reproduces the PQCD-predicted $1/\qsq$ shape 
of the transition form factor but disagrees with the absolute value given by Eqn.~\ref{EQ:41} 
by about~$15\%$ in the limit $\qsq \ra \infty$. 
The authors have stressed that this discrepancy 
is irrelevant in the region of $\qsq$ below 10 $\gevt$ and 
could, in principle, be eliminated by including the QCD evolution 
into the theoretical analysis~\cite{RR:comm_97}. 
It should be noted that the discussed theoretical analysis 
exactly reproduces the asymptotic prediction of 
PQCD given by Eqn.~\ref{EQ:41} when both photons are highly off-mass shell. 
We should emphasize that at present the non-perturbative treatment of various 
exclusive processes in a way similar to the approach of 
Radyushkin \etal ~is the subject of significant theoretical interest. 
For example, the QCD sum-rules method has been 
employed recently to predict the form factors in the semileptonic decays 
of the $B$ mesons\footnote{
Recent results of other theoretical developments 
relevant to our experimental study can be found in~the~literature~\cite{KROLL:WUB9619,MELIKHOV:97,DOROKHOV:95,SO:95,DAVIDSON:96,FRANK:95,KEKEZ:96,ANSELM:9603444,BELYAEV:9605279,SS:97}. 
}~\cite{BALL:93,ALIK:96,BRAUN:97}. 

The $\mgsg$ transition form factors 
have been studied by several experiments. 
The \mbox{LEPTON-G} experiment measured 
$\fetazg$ and $\fetapg$ 
in the time-like momentum transfer region 
up to 0.24 ~$\gevt$ 
using the rare electromagnetic decays 
$\etaz \ra \mu^+\mu^-\gamma$ and 
$\etap \ra \mu^+\mu^-\gamma$~\cite{LEPTONG:ff}. 
In order to achieve higher values of $\qsq$, the space-like 
photons produced in two-photon 
interactions were utilized 
by the PLUTO experiment to measure 
$\fetapg$ up to 1 $\gevt$~\cite{PLUTO:ff} 
and by the ${\rm TPC/2}\gamma$ collaboration to study 
$\fetazg$ and $\fetapg$ up to 7 $\gevt$~\cite{TPC:ff}. 
More recently, the CELLO experiment measured  
$\fpizg$ at $\qsq$ up to $2.7$ $\gevt$ and 
$\fetazg$ and $\fetapg$ at $\qsq$ up to $3.4$ $\gevt$~\cite{CELLO:ff}. 

We employ two-photon interactions 
to measure the transition form factors $\frg$ 
in the space-like regions of the momentum transfer 
between $1.5$ and $9$ $\gevt$ for $\piz$, 
$1.5$ and $20$ $\gevt$ for $\etaz$, 
and 
$1.5$ and $30$ $\gevt$ for $\eta^{\prime}$. 
We study the transition form factors 
of $\piz$, $\etaz$, and $\etap$ using the decays: 
\begin{center}
$\piz \ra \gamma\gamma$,
\end{center}
\begin{center}
$\etaz \ra \gamma\gamma$,
\end{center}
\begin{center}
$\etaz \ra \piz\piz\piz \ra 6\gamma$,
\end{center}
\begin{center}
$\etaz \ra \pip\pim\piz \ra \pip\pim2\gamma$,
\end{center}
\begin{center}
$\etap \ra \rhoz\gamma \ra \pip\pim\gamma$,
\end{center}
\begin{center}
$\etap \ra \pip\pim\etaz \ra \pip\pim2\gamma$, 
\end{center}
\begin{center}
$\etap \ra \piz\piz\etaz \ra 6\gamma$, 
\end{center}
\begin{center}
$\etap \ra \pip\pim\etaz \ra 2\pip2\pim2\gamma$, 
\end{center}
\begin{center}
$\etap \ra \piz\piz\etaz \ra 5\piz \ra 10\gamma$, 
\end{center}
\begin{center}
$\etap \ra \piz\piz\etaz \ra 3\piz\pip\pim \ra \pip\pim6\gamma$, 
\end{center}
\begin{center}
$\etap \ra \pip\pim\etaz \ra \pip\pim3\piz \ra \pip\pim6\gamma$.
\end{center}
We analyze the last two decay chains of $\etap$ together since 
they are observed in the same final state $\pip\pim6\gamma$. 

This paper is structured as follows: 
Section~II describes the CLEO~II detector 
and the data sample that we use for our measurements. 
Event selection criteria, experimental technique, and 
the analysis procedure for $\gaga$ final states 
are explained in Section~III. 
Analyses of other final states with 
only photons are described in Section~IV and 
analyses of final states with charged pions are described in Section~V. 
The unfolding procedure for the transition form factors is described in Section~VI. 
The results are compared  with some existing theoretical predictions in Section~VII. 
Conclusions are presented in Section~VIII. 

\section{Experimental Apparatus and Monte Carlo Simulation}

\subsection{The CLEO~II Detector and Data Sample}

The CLEO~II detector~\cite{CLEO-II:detector} is a general-purpose magnetic 
spectrometer which provides energy and momentum 
measurements for elementary particles. 
It is operated at the Cornell Electron Storage Ring 
(CESR), a symmetric $e^+e^-$ collider running at a 
center-of-mass energy near 10.6 GeV. 
The major objectives of the CLEO experiment are 
the studies of the properties of heavy mesons that contain $b$ or $c$ quarks. 
However, owing to the versatility of the detector, 
analyses of tagged and untagged two-photon interactions, 
detailed studies of $\tau$-lepton decays, 
and careful examination of quark and gluon fragmentation 
and other processes are also possible. 
 
The active components of CLEO~II include 
central tracking detectors, 
time-of-flight (TF) scintillator counters, 
muon detectors, and a CsI calorimeter for electromagnetic showers. 
The calorimeter consists 
of a barrel part covering polar angles above~$37^{\circ}$ 
and 
two endcap parts each covering the region between~$13^{\circ}$ and~$37^{\circ}$, 
where the polar angle is measured with respect to the beam axis. 
The energy resolution of the barrel calorimeter 
for photons of energies above 500 MeV is~2\%. 
The central tracking detectors consist of 
three concentric cylindrical drift chambers that 
cover the polar angles above~$18^{\circ}$. 
From smallest to largest radii these are: the precision 
tracking layers detector, the vertex detector (VD), and 
the main drift chamber. 
The measurements of the 
specific ionization energy losses 
in the outer layers of the main drift chamber 
and 
flight times 
in the TF system 
provide discrimination between 
charged particles of different species. 
All detector subsystems except the muon detectors 
reside in a uniform axial magnetic field 
of 1.5 Tesla. 

The data sample employed in our analysis 
corresponds to an integrated $e^+e^-$ luminosity 
of $2.88 \pm 0.03$ fb$^{-1}$. 
Two thirds of the data was collected 
at $e^+e^-$ center-of-mass energy of $\sqrt{s} = 2 E_b = 10.58$ GeV, 
the remainder at 10.52 GeV. 

\subsection{Trigger System}

The CLEO~II detector has a three-level hardware trigger 
system~\cite{CLEO-II:trigger} 
followed by a software filter.  The fastest, ``zeroth'' level (L0) trigger 
can be either track-based (using the VD and TF) or energy-based 
(demanding a minimum energy deposition of about 500 MeV in the CsI calorimeter). 
The calorimeter L0 information 
develops slowly, so it only comes into effect if the track-based L0 trigger fails; 
in such cases the tracking information is lost. 

The first level (L1) trigger uses track-based information from 
the VD, TF, and main drift chamber; tracks of transverse momenta in excess of 
about~340 MeV/c are identified by either of two independent track processors 
employed in the trigger decisions. 
Calorimeter information is also utilized at L1. High threshold bits, 
designed to be set by showering particles, have a threshold of about $500$ MeV; 
low threshold bits, designed to trigger on minimum ionizing particles, 
have a threshold of about 100 MeV.  
To trigger at L1 on two low-energy clusters, 
they must be well-separated in space. 

More detailed information from the VD and main 
drift chamber is used in the second level (L2) trigger. 
The requirements and accessed momentum range varied between 
data subsets, but are all modeled in our detector simulations. 
The software filter (LVL3) is optimized to suppress backgrounds from 
interactions of the beams with residual gas and vacuum chamber walls. 
Events which pass LVL3 are recorded. 
In addition, every eighth event that fails LVL3 is also recorded 
to allow the LVL3 efficiency to be studied. 

The efficiencies of the various trigger components have been measured
using data collected with independent or partially independent simultaneous
trigger requirements 
and are incorporated in the detector simulations~\cite{SAVINOV:thesis,ACOSTA:thesis,ONG:thesis}. 
The simulations are carefully run to match the integrated luminosity 
associated with each trigger configuration. This is necessary 
because exact trigger requirements in CLEO~II have been changed over time 
to improve the trigger efficiency for events of low particle multiplicities. 

\subsection{Monte Carlo Simulation}

In our analysis we use a two-photon Monte Carlo (MC) simulation program 
\cite{COFFMAN:MC} 
that is based on the BGMS formalism~\cite{BGMS:75}. 
The $\mgsgs$ transition form factors are approximated by: 
\begin{eqnarray}
%\[
|\frgs(Q^2,q^2)|^2 & = & \,
|\frg(Q^2)|^2 \,
\frac{1}{(1+q^2/\Lambda_{\cal R}^2)^2} \,
\, \nonumber
\\ & = &
\frac{1}{(4 \pi \alpha)^2} \,
\frac{64 \pi \Gamma(\cal R \ra \gaga)}{M_{\cal R}^3} \,
\frac{1}{(1+Q^2/\Lambda_{\cal R}^2)^2} \,
\frac{1}{(1+q^2/\Lambda_{\cal R}^2)^2} \,
, 
%\]
\label{EQ:137}
\end{eqnarray} 
where $Q^2$ and $q^2$ are the absolute values of the squared 
four-momenta carried by the space-like photons. 
The pole-mass parameter \mbox{$\Lambda_{\cal R}$ = 770 MeV} has been chosen 
to approximate the momentum transfer dependence of the form factors. 
It should be noted that while we have chosen this parameter 
to be practically the $\rho^0$ mass, 
as predicted by the Vector Meson Dominance (VMD) model~\cite{SAKURAI:60},  
the pole-mass behavior of the transition form factors $\frg$ 
and the value of the parameter $\Lambda_{\cal R}$ 
in the range between 700 and 900 MeV are indicated by 
various theoretical predictions~\cite{KROLL:96,RR:9603408} 
that are not based on VMD. 
Notice that in the approximation given by Eqn.~\ref{EQ:137} we assume a factorization 
of the form factor into the $q^2$- and $Q^2$-dependent parts~\cite{GRIBOV:62}. 
In the same two-photon simulation program 
we also generate the decays of the produced mesons. 
To account for 
the relativistic effects, 
helicity conservation, 
and 
presence of spin-one particles 
we simulate the decay chain $\etap \ra \rhoz\gamma \ra \pip\pim\gamma$ 
according to: 
%~\cite{COFFMAN:THANKS}: 
\begin{equation}
\frac{d^2\Gamma(\etap \ra \rhoz\gamma \ra \pip\pim\gamma)}
     {d\cos{\theta^*}\,dm^2_{\pi\pi}} 
\propto 
\sin^2{\theta^*} \, 
\frac{{E}^3_{\gamma}}{m_{\pi\pi}} \, 
\frac{m_{\rho} \Gamma(m_{\pi\pi})}
{(m^2_{\rho}-m^2_{\pi\pi})^2+m^2_{\rho} \Gamma^2(m_{\pi\pi})},
\label{EQ:EPRG1}
\end{equation}

\noindent with the energy-dependent width, 
$\Gamma(m_{\pi\pi})$, parameterized by: 

\begin{equation}
\Gamma(m_{\pi\pi}) = \Gamma(m_{\rho}) \, 
\frac{|\vec{p_a}|^3}{|\vec{p_n}|^3}, 
\label{EQ:EPRG2}
\end{equation}

\noindent where $\theta^*$ is the angle between the directions 
of one of the charged pions and the signal photon, 
$E_{\gamma}$ is the energy of the photon, 
$\Gamma(m_{\rho}) = 151$ MeV and $m_{\rho}=768$ MeV/c$^2$ are the nominal 
width and mass of $\rhoz$~\cite{PDG:96}, 
$m_{\pi\pi}$ is the actual mass of $\rhoz$, 
and 
$|\vec{p_a}|$ and $|\vec{p_n}|$ 
are the magnitudes of the charged-pion momenta 
for the actual and nominal masses of $\rhoz$, 
respectively. 
The charged-pion momenta and the angle $\theta^*$ 
are defined in the center-of-mass frame of $\rhoz$ 
and 
the energy of the photon 
is defined in the center-of-mass frame of $\etap$. 

The transport of the generated MC particles through the CLEO~II detector 
is performed by a GEANT-based~\cite{CERN:GEANT} detector 
simulation program.  
The generated events are then processed 
by the event reconstruction program 
which also ``simulates'' 
random electronic noise and beam-related spurious energy clusters 
by adding hits from random-trigger data samples 
into the MC events. 

%-----------------Start pi0 analysis & results--------------------------

\section{Analyses of Single-Tagged $\gaga$ Final States}

\subsection{Trigger}

The single-tagged two-photon reactions 
$e^+e^- \ra e^+e^-\piz$  and 
$e^+e^- \ra e^+e^-\etaz$ 
followed by the decays $\pi^{0}$$\ra$$\gamma\gamma$ 
and $\eta$$\ra$$\gamma\gamma$ 
are recorded using either a track-based or an energy-based L0 trigger. 
The track-based L0 trigger is satisfied when the scattered electron 
passes through the VD and enters the endcap TF. 
For tags that scatter at polar angles above~$24.5^{\circ}$, 
thus passing through the entire VD volume, 
the efficiency of this trigger is about~$80\%$ and 
is determined by the
size of the wire-chamber drift cells compared to 
the time allowed to make the L0 decision. 
At smaller polar angles we rely on the energy-based L0 trigger. 
The efficiency of this trigger is~98\%~(100\%) for  
electrons which deposit~1.0 GeV (more than~1.6 GeV) of energy in the calorimeter. 

The L1 trigger 
is satisfied when 
at least two clusters, each of energy above 500 MeV, 
are detected in the calorimeter, with
one in the barrel region and the other in one of the
endcap regions.
There are no L2 requirements for events passing
the L0 and L1 trigger conditions described above. 

To be recorded, events must fulfill the 
transverse-momentum requirement of the LVL3 filter 
that assigns momenta to all calorimeter 
clusters assuming that they are photons 
produced at the primary interaction point in the geometrical 
center of the CLEO~II detector. 
This LVL3 criterion rejects events if the
net vector momentum has a component
normal to the beam axis in excess of
0.7 GeV/c (1.4 GeV/c) when the total energy detected in 
the calorimeter is larger than 1.0 (5.0) GeV. 

\subsection{Analysis Procedure}

In the first part of this section we describe 
the event selection criteria based on the event topology 
for the signal production processes. 
In the second part we explain selection criteria aimed 
at the suppression of random background. 
In the third part we discuss the event quality requirements 
designed to isolate signal events with large uncertainty in the 
detection efficiency. 
Finally, in the last part of this section 
we show the invariant mass spectra for data events that 
fulfill all selection criteria. 

\subsubsection{Basic Selection Criteria}

The event selection criteria for single-tagged $\gaga$ final states 
are designed to isolate two-photon events 
for which the trigger efficiency is high 
and 
in which the only missing particle is the untagged electron 
of high momentum. 
These events are characterized by the 
high-energy shower produced by the tag in the endcap calorimeter 
and 
two electromagnetic showers of total energy larger than 1 GeV 
produced by the photons in the barrel calorimeter. 

We select events in which 
three or four energy clusters 
and no more than one charged track 
have been reconstructed. 
The energy of each barrel (endcap) cluster must be larger than 30 (50) MeV. 
The most energetic cluster is assumed to be produced by the tag and 
must be in the endcap calorimeter. 
If a charged track is found, 
its projected intersection point with the calorimeter must agree 
with the tag's shower position 
within~$20^{\circ}$ as estimated at the primary interaction point. 
The position of each shower is 
determined from the energy-weighted average of the 
centers of the crystals forming this shower. 
To provide an efficient trigger, the energy of the tag candidate 
detected in the calorimeter should be above~1.0~GeV 
(at a later stage of the analysis procedure 
this cut will be superseded by a tighter requirement). 
Out of the remaining energy clusters, the two most energetic 
must be found in the barrel calorimeter at polar angles 
above~$45^{\circ}$ ({\it i.e.}, excluding calorimeter edges), 
and are assumed to have come from the $\piz$ or $\etaz$ decays. 
The fourth energy cluster, if found, should contain less 
than 200 MeV of energy; 
the efficiency loss due to this requirement is less than~0.25\%. 
Events with this additional energy cluster may be either signal or beam-gas 
events with a beam-related noise cluster or partially reconstructed background 
events of higher particle multiplicities that mimic 
single-tagged $\piz$ or $\etaz$ production. 
By allowing an extra energy cluster, we 
reduce the uncertainty in the signal efficiency while providing 
the opportunity for background estimates. 
A tighter cut on the energy of an additional cluster would make our results 
more sensitive to the modelling of the noise-related energy clusters 
and 
a looser cut on this extra energy would not adequately discriminate 
against signal-like background which is due to partially reconstructed events. 

The overall efficiencies of the basic selection criteria described above 
are~38\% and~30\% for the $\piz$ and $\etaz$ analyses, respectively. 
These estimates have been obtained using MC signal events 
generated in the $\qsq$ range between~1.5 and~9 $\gevt$. 

\subsubsection{Background Suppression}

The background conditions 
in two-photon events of low particle multiplicities 
with tags detected at large and (relatively) small 
polar angles are different. 
To provide an adequate background suppression 
for both regions of polar angle, 
we separate signal event candidates into two samples 
that have undergone different experimental cuts. 
In this subsection we describe 
this event separation, 
the sources of random background 
and 
the event selection criteria applied to each sample 
to suppress random background. 

When the scattering angle of the tag 
is larger than~$24.5^{\circ}$ (as determined from the calorimeter) 
we select events that have been triggered by the track-based L0 trigger. 
In addition, we require that these events have exactly one reconstructed charged 
track consistent with the tag's shower. 
There is no efficiency loss associated with the tracking requirement 
which discriminates against background arising mainly from radiative Bhabha 
events accompanied by photon conversion or bremsstrahlung. 
We include these events in the track-tagged sample. 
When the scattering angle is less than~$24.5^{\circ}$ 
we accept both track- and energy-based L0 triggers 
and do not require the presence of the tag's track, 
because 
the efficiencies of the 
track-based L0 trigger 
and 
track reconstruction 
vanish for tags detected in this region of polar angles. 
We include these events 
that have been triggered either 
by the track-based 
or 
energy-based L0 trigger 
in the energy-tagged sample. 
Notice that while 
the events from the track-tagged  sample must  be track-triggered, 
the events from the energy-tagged sample could be either track- or energy-triggered. 
Tracking information for events from the track-tagged sample 
is utilized in background estimates. 
The track reconstruction efficiency for energy-triggered events is zero. 

Before imposing further selection criteria we obtain 
improved estimates of the tag energy and direction by 
using transverse-momentum balance and 
the tag coordinates in the calorimeter. 
The transverse momenta of the tag and of the photon pair 
should be nearly identical for signal events 
because the untagged electron usually 
carries very little transverse momentum (below 5~MeV/c) 
according to the prediction of the MC simulation. 
Since the transverse momentum of the photon pair is 
measured with much better precision than that of the tag, 
we equate the magnitude of the transverse momentum of the tag 
with that of the signal photon pair. 
To calculate the direction of the tag we require that 
its trajectory in the magnetic field goes through the center of the tag's shower. 
To estimate the center of the shower we use 
the measurement from the calorimeter 
when the shower is found at polar angles larger than~$16.5^{\circ}$. 
At smaller polar angles, however, 
we use the geometrical center of the crystal with the largest detected energy. 
This is necessary in order to reduce the discrepancies between the data and MC simulation. 
Using the estimates of the tag energy and direction 
obtained from the transverse-momentum balance 
we estimate the missing energy and the magnitude of missing momentum. 
To suppress the background from partially reconstructed events 
we select events where the discrepancy between the missing energy and missing momentum 
is less than~2.3 GeV. This cut is~98\% efficient for our signal. 
We note that at a later stage of the analysis procedure 
we will obtain more precise estimates of the tag energy and direction. 
 
Not only should the magnitudes of the transverse momenta of the tag and the photon pair 
be nearly equal, their directions 
are expected to be practically opposite 
in the plane perpendicular to the beam collision axis. 
We use the acoplanarity angle, 
which is the deviation from this expectation, 
to suppress the background 
arising from radiative Bhabha events with bremsstrahlung photons 
produced in the materials of the detector. 
An event of this origin enters the energy-tagged sample 
when the track-based L0 trigger is 
inefficient and a track associated with 
an electron which radiated in the barrel part of the detector 
cannot be reconstructed. 
While for signal events the acoplanarity distribution peaks near zero, 
for background events it peaks around ~$12^{\circ}$ for the CLEO geometry 
and CESR kinematics. 
Acoplanarity discriminates between signal and background events 
because the measured angular position of the shower created by the electron 
that has undergone bremsstrahlung 
is shifted with respect to its direction at the primary interaction point. 
This shift is due to the bending of the electron track in the magnetic field. 
To suppress this random QED background in the energy-tagged sample, 
we select events with acoplanarity less than~$5^{\circ}$. 
The background rejection power of this cut exceeds~10, 
while efficiency loss varies 
between~20\% and~10\% for $\qsq$ 
between~1.5 and~2.5 $\gevt$. 
For $\qsq$ larger than~2.5 $\gevt$ 
the efficiency of the acoplanarity cut 
for the energy-tagged events is~90\%. 
In contrast to the energy-tagged sample, the track-tagged sample 
contains very few bremsstrahlung-accompanied radiative Bhabha events 
because each of these background events has an additional charged track 
and does not pass basic selection criteria 
(the track reconstruction efficiency for high-energy electrons 
detected in the barrel part of the detector is practically~100\% 
for events recorded by the track-based L0 trigger). 
We select the track-tagged events with acoplanarity 
less than~$15^{\circ}$. 
The efficiency of this loose cut on acoplanarity is~99\%. 

We use the decay angle $\theta_d$ to further suppress background 
arising from radiative Bhabha events accompanied by 
low-energy split-off clusters. 
The decay angle is determined from 
the directions of the $\piz$ (or $\etaz$) candidate in the lab frame 
and 
one of the daughter photons in the center-of-mass frame of $\piz$ (or $\etaz$). 
Simulation of the detector acceptance 
predicts that the distribution of $|\cos{\theta_d}|$ 
is flat between~0.0 to~0.95 and decreases rapidly 
beyond~0.95 due to the acceptance loss for soft photons. 
In contrast to the signal, radiative Bhabha events 
with split-off clusters 
congregate at $|\cos{\theta_d}| = 1.0$ because these clusters 
typically are of low-energy. 
We reject these asymmetric decays by requiring 
$|\cos{\theta_d}| < 0.90$. 

The acoplanarity and decay angle cuts do not eliminate 
random background completely, 
because radiative Bhabha events accompanied by 
$\gamma$ conversions in detector materials 
look similar to signal events 
when triggered by the energy-based L0 trigger. 
However, we have found that the shape of this background 
is monotonic within the signal and sideband regions 
of the $\gaga$ invariant mass distribution in both analyses. 

\subsubsection{Event Quality Requirements}

The angular spectrum of the scattered electrons 
peaks sharply at small polar angles 
due to the kinematics of processes studied in our analyses. 
Thus, to measure the cross sections for two-photon production 
in a tagged mode we must understand this critical region 
of our experimental apparatus very well. 
While we can, in principle, 
detect tags at polar angles as small as~$13^{\circ}$, 
the fraction of the tag energy collected in the calorimeter 
at these small polar angles is usually less than~20\% and 
might be insufficient to trigger an event. 
In addition, even if the trigger is satisfied, 
an event might be rejected by the LVL3 filter, 
which is biased against events with large net transverse momenta. 
To select events identified in the detector regions 
where the trigger and LVL3 efficiencies are well understood, 
we need better estimates of the tag energy and scattering angle. 

\begin{figure}
\centerline{
	\psfig{figure=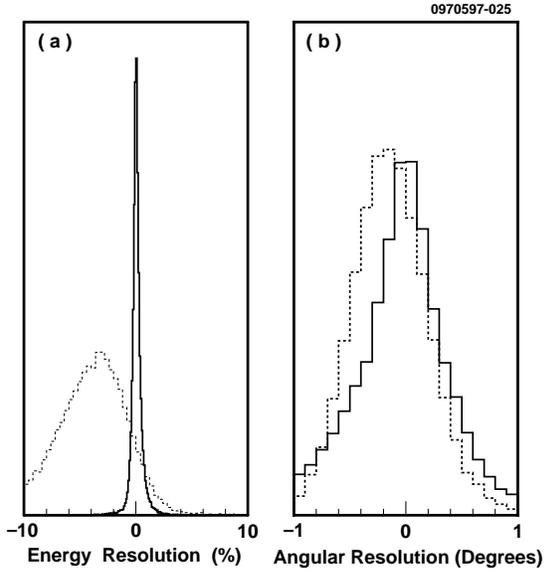,height=3.00in}
           }
\caption{
Resolution functions of a) energy (in \%) and b) scattering angle (in degrees) 
obtained from MC simulation in the $\piz$ analysis. 
Dashed and solid lines show resolution functions 
measured directly in the calorimeter 
and 
achieved using energy-momentum conservation, 
respectively. 
}
\label{fig:fig_01}
\end{figure}

To make precise estimates of the tag energy and scattering angle 
we use energy-momentum conservation assuming that 
the only particle missing detection 
is the untagged electron with zero transverse momentum. 
In practice, this method allows us to estimate the 
parameters of the tag when we 
measure only the four-momentum of the hadronic system 
and 
assume that we know the charge of the untagged electron 
(from crude measurement of the direction of missing momentum). 
From conservation laws we estimate 
the tag energy $E$ with an r.m.s. resolution of~0.003~$E$ 
and 
the scattering angle with an r.m.s. resolution of better than~$0.6^{\circ}$. 
In addition, to estimate the scattering angle 
for track-tagged events we use 
the polar angle of the reconstructed charged track 
associated with the tag. 
By using the polar angle of the track we achieve an additional 
small improvement in the resolution of the scattering angle 
for these events. 
The tag energy for track-tagged events, however, is estimated from 
energy-momentum conservation; {\it i.e.} no tracking information 
is used to estimate the tag energy. 
In further discussions the values of the tag's parameters estimated 
from energy-momentum conservation 
and the polar angle of the charged track are referred to 
as constrained values of the tag energy and scattering angle. 
In Fig.~\ref{fig:fig_01} 
we show the resolution functions of the 
tag energy and scattering angle 
determined from the differences 
between analyzed ({\it i.e.} measured or constrained) and generated quantities 
(normalized to the generated value for the energy resolution function). 
These resolution functions have been obtained using simulated $\piz$ events 
which have values of $\qsq$ between~1.5 and~9 $\gevt$ and 
satisfy all selection criteria discussed above. 
In our analyses we estimate $\qsq$ for each event using 
constrained values of the tag energy and scattering angle. 
This results in an r.m.s. $\qsq$ resolution that varies between 
0.1 and 0.3 $\gevt$ for the $\qsq$ region between 
1.5 and 9   $\gevt$. 

\begin{figure}
\centerline{
	\psfig{figure=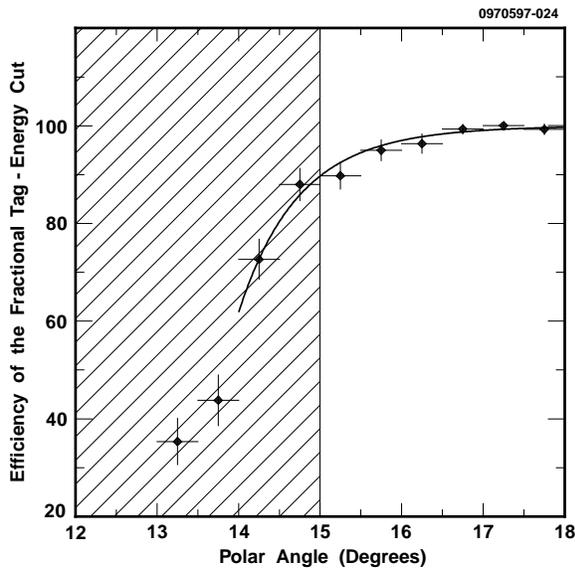,height=3.00in}
           }
\caption{
The efficiency (in \%) of the fractional tag-energy cut as measured from data. 
The solid line shows a power law approximation chosen 
to interpolate between the efficiency measurements. 
Events with tags scattered at polar angles less than~$15^{\circ}$ are rejected from all analyses. 
}
\label{fig:fig_02}
\end{figure}

To isolate the detector region for which the efficiency is small and 
poorly understood, 
events with constrained values of the tag scattering angle less than~$15^{\circ}$ 
are rejected from further analysis. 
In addition, to reduce the systematic uncertainty 
in the efficiency of the LVL3 filter 
we select events in which the detected fraction of the tag energy is at least~50\%. 
This fraction is estimated 
from the calorimeter measurement and the constrained value of the tag energy. 
The efficiency of this fractional tag-energy cut is~90\% for 
tags which scatter at~$15^{\circ}$ and 
is practically~100\% for tags which scatter at angles larger than~$19^{\circ}$. 
We have measured the dependence of this efficiency 
on the polar angle using radiative Bhabha events triggered 
inclusively by the barrel TF-based L0 trigger. 
We show the efficiency of the fractional tag-energy cut 
in Fig.~\ref{fig:fig_02}. 

\subsubsection{Event Selection Results}

\begin{figure}
\centerline{
	\psfig{figure=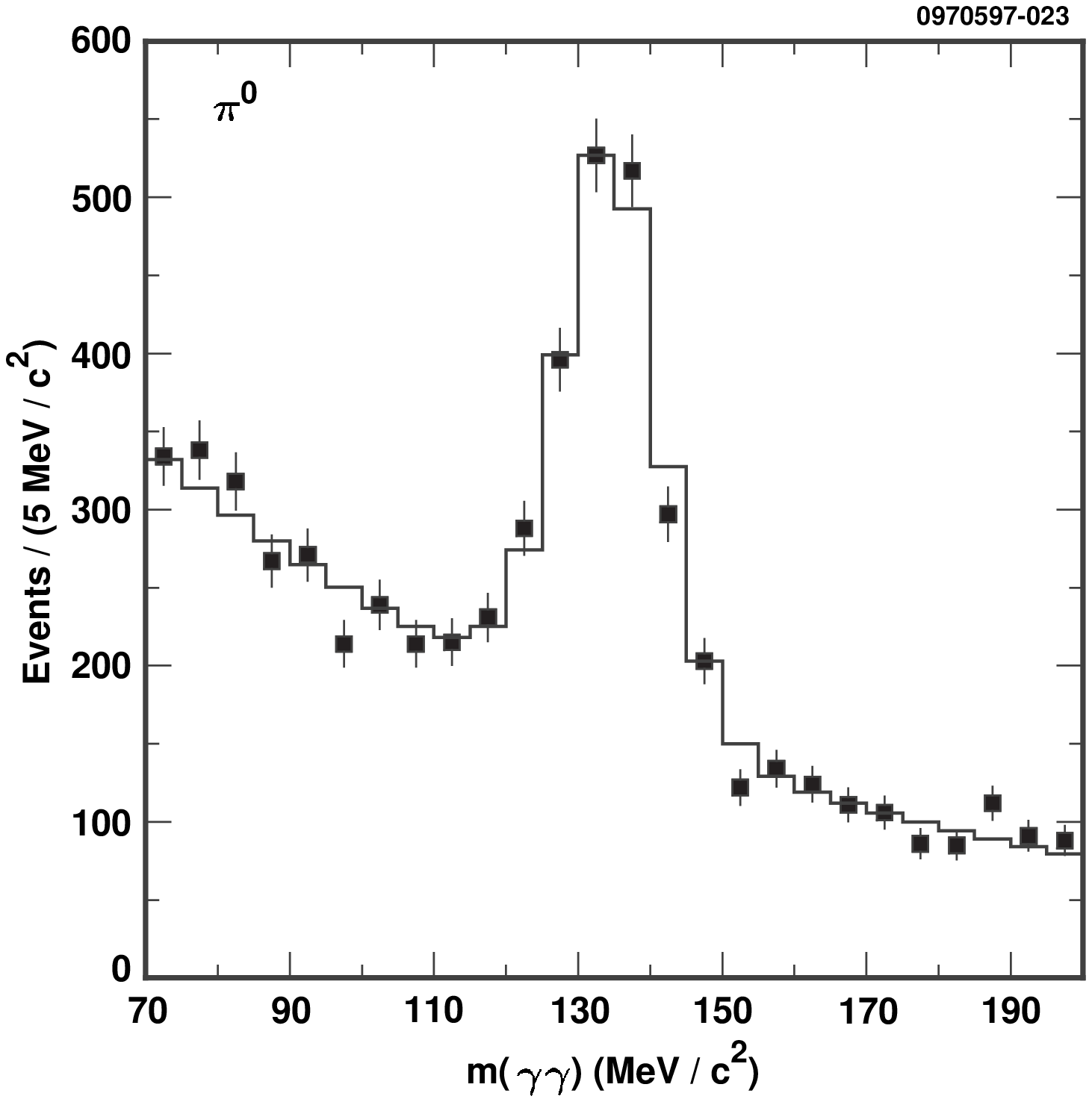,height=3.00in}
           }
\caption{
Fit (solid line) to the $\gaga$ invariant mass distribution 
observed in data (points with error bars) in the $\piz \ra \gaga$ analysis. 
The signal line shape is obtained from the MC simulation, 
the remaining random background is approximated by an exponential. 
}
\label{fig:fig_03}
\end{figure}

\begin{figure}
\centerline{
	\psfig{figure=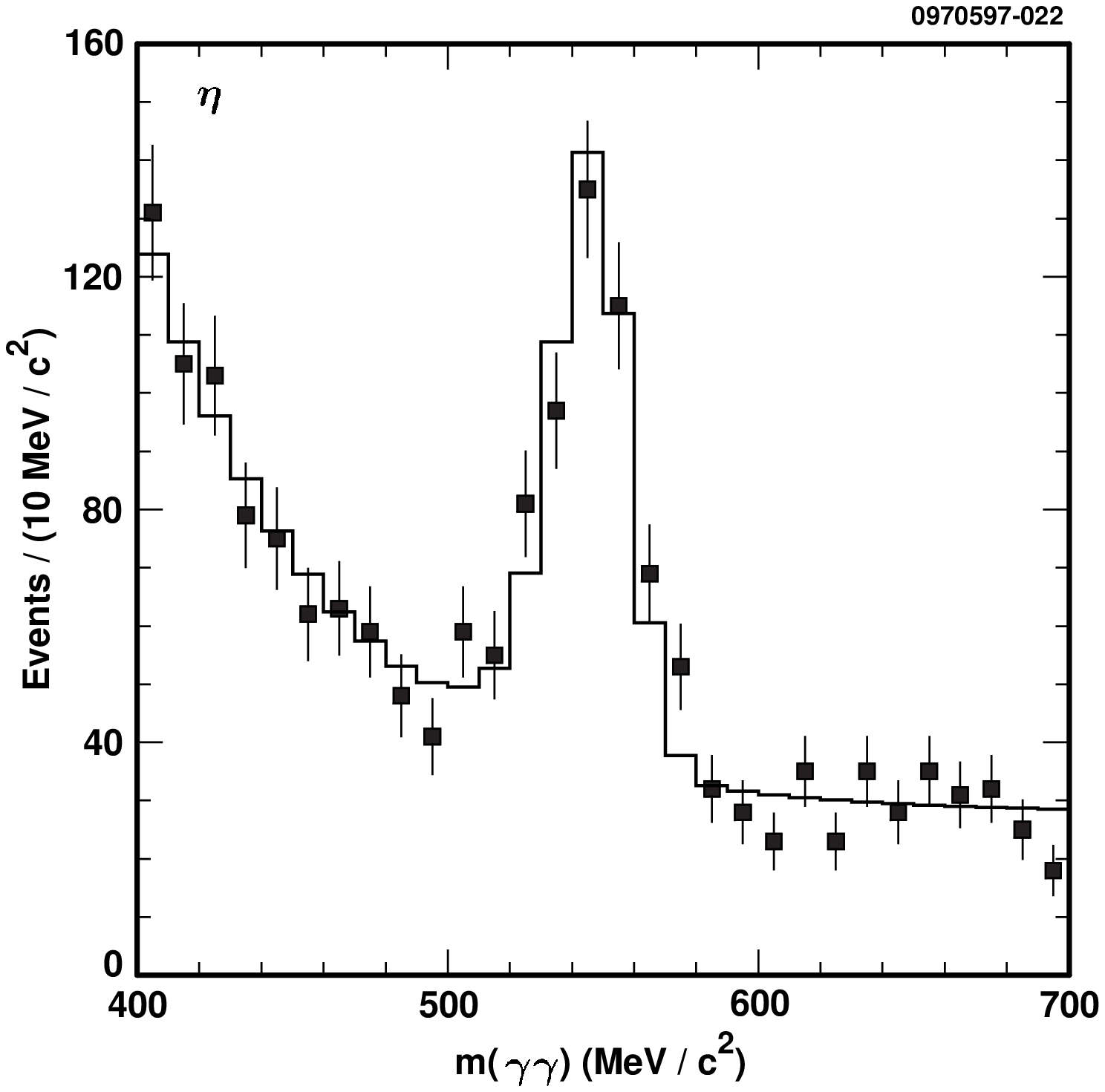,height=3.00in}
           }
\caption{
Fit (solid line) to the $\gaga$ invariant mass distribution 
observed in data (points with error bars) in the $\etaz \ra \gaga$ analysis. 
The signal line shape is obtained from the MC simulation, 
the remaining random background is approximated by 
the sum of an exponential and a constant. 
}
\label{fig:fig_04}
\end{figure}

In Figs.~\ref{fig:fig_03}~and~\ref{fig:fig_04} 
we show the $\gaga$ invariant mass distributions 
for data events that pass all selection criteria 
for the $\piz$ and $\etaz$ candidates 
and have values of $\qsq$ between 1.5 and 9 $\gevt$. 
The points with error bars in these figures 
represent event yields in data. 
The solid line in each figure shows the result of the binned likelihood fit 
to data with the signal line shape obtained from the MC simulation 
and an approximation of the remaining random background. 
In the $\piz \ra \gaga$ analysis, the background arising from 
radiative Bhabha events accompanied by photon conversions 
is approximated by an exponential. 
In the $\etaz \ra \gaga$ analysis 
random background is approximated by the sum of an exponential and a constant 
because the $\gaga$-mass distribution shown in Fig.~\ref{fig:fig_04} 
contains two major background components. 
While the first component has the same source as in the $\piz$ analysis, the second 
component is due to radiative Bhabha events with bremsstrahlung radiation in 
the interface between the drift chambers. 

\subsection{Background Estimates}

The data may contain $\piz$ and $\etaz$ events 
that are due to 
beam-gas interactions 
or 
partially reconstructed events of higher particle multiplicities. 
To estimate the beam-gas contribution 
we use 
the distributions of 
the event vertex position, 
visible energy 
and 
squared missing mass. 
Given the profile of the residual gas density near the 
beam-collision point, the vertex position of beam-gas events 
is much more diffuse than that of the signal. 
In addition, while beam-gas events should have 
visible energy ({\it i.e.}, total energy detected in the calorimeter) 
less than the beam energy, 
most events from the track-tagged sample have visible energy larger than 
the beam energy. 
However, 
at small scattering angles the tag needs to go through 
a larger amount of the detector materials than at large scattering angles 
and can lose a significant part of its energy 
before reaching the calorimeter. 
As a result, a large fraction of events from the energy-tagged sample (about~20\%) 
falls into the visible energy region below the beam energy. 
For these energy-tagged events we have studied the distribution of the squared missing mass 
estimated assuming the electroproduction hypothesis 
$e^{\pm} p \ra e^{\pm} p \piz ({\rm~or~}\etaz)$. 
Using the discriminating power of the distributions described above 
we conclude that the beam-gas background is very small 
and warrants no subtraction. 

To estimate the background contribution to the track-tagged sample 
due to $e^+e^-$ annihilation 
we have studied the correlation between the charge 
and the direction of the tag's track. 
Signal processes should produce virtually all positrons 
in the $+z$ hemisphere and electrons in $-z$ hemisphere, 
where $+z$ is the direction of the positron beam. 
However, 
$e^+e^-$ annihilation should produce practically 
the same number of electrons (and positrons) 
in both $z$-hemispheres. 
We do not observe a single data event 
in which this charge-direction correlation 
indicates $e^+e^-$ annihilation processes. 
We conclude that the background from $e^+e^-$ annihilation 
is fewer than 1 event in both track- and energy-tagged samples 
because the angular distribution of the electrons from this 
background source is expected to be relatively uniform 
(compared to the rapidly changing signal). 

Finally, there may be some background from other single-tagged 
two-photon processes. 
The process \mbox{$e^+e^- \ra e^+e^-f_2(1270)$} 
followed by the decay \mbox{$f_2(1270) \ra \piz\piz$} 
is the most likely source of the feed-down for the \mbox{$\piz \ra \gaga$} analysis. 
To estimate the feed-down from this process, 
we remove the cut on the energy of the fourth, 
least energetic cluster and repeat the analysis. 
We estimate that out of 1300 $\piz$ event candidates in data, 
$80 \pm 40$ events are due to the feed-down, where the error reflects 
the uncertainty of our method. 
This uncertainty arises from the fact that 
the $\piz$ misidentification probability 
for the feed-down from the decay \mbox{$f_2(1270) \ra \piz\piz$} depends 
on the relative strengths of the couplings between the tensor meson and 
two space-like photons of various total helicity 
(determined in the center-of-mass frame of $f_2(1270)$). 
The central value of the background estimate quoted above has been 
derived assuming that the $f_2(1270)$ production proceeds 
exclusively via the helicity $\pm2$ channel. 
The error reflects the uncertainty in the background estimate 
which becomes larger (smaller) 
when we assume that $f_2(1270)$ is produced only in the 
helicity 0 ($\pm1$) state. 
We assign this large error to the background estimate 
because 
the contributions of different helicity amplitudes 
to the single-tagged cross section for this background process 
have not been measured yet. 
We observe the $f_2(1270)$ feed-down at $\qsq$ below~4 $\gevt$ and subtract 
its contribution to each $\qsq$ interval 
using the shapes of the energy spectra of an additional cluster measured 
from data and signal MC simulation. 
We do not observe a feed-down in the \mbox{$\etaz \ra \gaga$} analysis. 
We have also studied the feed-down from single-tagged two-photon processes 
of higher final-state particle multiplicities 
such as the production of $\etaz$ and $\etap$ and estimate 
the overall contribution from these background processes 
to be insignificant in both analyses. 

\subsection{Systematics}

Contributions to the systematic errors arise from four sources. 
The primary uncertainty is due to systematic biases 
in the determination of the event selection efficiency. 
These biases are detailed below. 
The second contribution is a~1\% systematic 
error on integrated luminosity~\cite{BKH:luminosity}. 
This error is based on estimates 
of the theoretical uncertainties in the QED radiative corrections in 
the MC event generators for the processes 
$e^+e^- \ra e^+e^-$ and $e^+e^- \ra \gaga$ 
which are employed in the determination of 
integrated luminosity. 
The third contribution is a~1\% 
systematic error due to the background estimation procedure. 
The fourth source of systematic error is due to 
small uncertainties in the branching fractions for studied decay chains. 
This error 
is negligible in the $\piz \ra \gaga$ analysis 
and 
is less than~1\% in the $\etaz \ra \gaga$ analysis. 

The largest systematic error is 
due to the fractional tag-energy cut. 
We have measured the efficiency of this cut 
using radiative Bhabha events in data. 
The relative statistical error in this efficiency is less than~3\% 
for polar angles larger than~$15^{\circ}$ 
so we conservatively include a~3\% error to the systematics 
of energy-tagged events. 
Note that the fractional tag-energy cut is fully efficient for track-tagged events, 
so no contribution is made to their systematics. 

The efficiency of the LVL3 filter has been measured 
using $\piz$ signal data events that would 
have normally been discarded by this filter. 
The statistical error in the measured efficiency 
is~2\% and this gives an estimate of the systematic error. 

The next error comes from the uncertainty in the 
photon reconstruction efficiency. 
We have determined this uncertainty to be~2\%, 
or~1\% per photon from a global fit of the measured ratios 
of the $\etaz$ and $\etap$ branching fractions 
to their average values~\cite{PDG:96}. 

We have measured the efficiency of the VD L0 trigger 
over the entire data sample using the TF-triggered 
endcap Bhabha events and have found that this efficiency 
varies by up to~2.5\% of its central value between data subsets. 
In our analysis we use the average value for the VD L0 trigger efficiency of~80\% 
and include its \mbox{r.m.s.} variation 
of~2\% to the systematic error for track-tagged events. 

To estimate the systematic uncertainty in the 
efficiency of the extra energy cut 
we have utilized the shape of the extra energy distribution 
measured from signal data 
in the $\etap \ra 6\gamma$ analysis. 
We estimate this uncertainty to be~2\%. 

The efficiency of the acoplanarity cut for energy-tagged events 
is between~80\% and~90\%, depending on $\qsq$. 
To estimate the uncertainty in this cut, 
we have measured its efficiency 
assuming that the detector simulation systematically 
underestimates or overestimates azimuthal angular 
positions of all showers and the tag in the calorimeter by one standard deviation 
of the angular resolution function. 
We find that under these conditions 
the efficiency varies by less than~1\% of itself 
in any $\qsq$ interval. 
We include this value of~1\% to the systematics of energy-tagged events. 

We have also studied other sources of uncertainties such as 
the efficiencies of missing energy-momentum and decay angle cuts 
and conclude that their total contribution to the systematics 
is insignificant. 

We include the systematic uncertainties 
in the amount of feed-down background and 
in the shape of the $\gaga$-mass spectrum for random background 
to the statistical error on the number of signal events 
in each $\qsq$ interval. 
These errors are between~1\% and~5\% being larger at smaller $\qsq$. 

\begin{figure}
\centerline{
	\psfig{figure=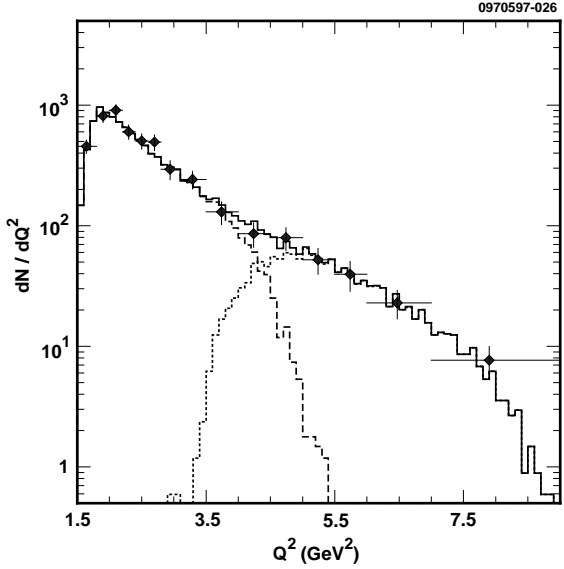,height=3.00in}
           }
\caption{
The $\qsq$ distributions for signal $\piz$ events 
in MC (solid line) and data (points with error bars) 
in the $\piz \ra \gaga$ analysis. 
The distributions for events which belong 
to the energy- and track-tagged MC samples 
are shown with dashed and dotted lines, 
respectively. 
For each $\qsq$ interval in data the number of signal events 
is obtained from the fit followed by the background subtraction. 
The number of MC events is normalized to the number of signal $\piz$ events in data. 
}
\label{fig:fig_05}
\end{figure}

While the acoplanarity and fractional tag-energy cuts affect only energy-tagged events, 
the track-based L0 trigger is specific for the track-tagged events. 
Thus, the systematic uncertainties associated with the two event samples are different. 
To estimate the systematics for each $\qsq$ interval 
we have used the $\qsq$ distributions for MC events which belong 
to the energy- and track-tagged samples. 
We show these distributions in Fig.~\ref{fig:fig_05}. 

Our analyses should not be 
significantly affected by the QED radiative corrections. 
To order $\alpha^5$, 
in addition to the vacuum polarization 
and 
one virtual photon exchange, 
these corrections 
describe the processes $e^+e^- \ra e^+e^-{\cal R}\gamma$, where ${\cal R}$ 
is one of the studied pseudoscalar mesons~\cite{RC:NIKHEF}. 
When a radiative photon carries away part of the 
initial center-of-mass energy and remains undetected, 
we use the nominal value of the beam energy before the radiation 
and overestimate~$\qsq$ according to Eqn.~\ref{EQ:12}. 
However, when we estimate the tag energy and scattering angle 
from energy-momentum conservation, we underestimate~$\qsq$. 
Both distortions described above are small effects because 
the energy spectrum of radiative photons is very soft. 
We neglect the effect of the QED radiative corrections 
on the smearing of the $\qsq$ spectrum 
because 
these two small effects largely cancel each other. 
The net smearing is such that in our analysis procedure 
the measured cross sections are insignificantly underestimated. 

There is another aspect of the QED radiative corrections 
that might need to be taken into account. 
Namely, when we unfold the differential cross sections 
and obtain the transition form factors, 
we rely on the prediction of a numerical integration 
that does not contain these corrections and underestimates the cross sections. 
We expect the QED radiative corrections to the cross sections 
for single-tagged events to be smaller than~2.4\%~\cite{RC:NIKHEF,RC:TRONDHEIM} 
and this gives a~1.2\% estimate of the systematic uncertainty 
introduced in the values of $\frg$ from the unfolding procedure\footnote{
If we include the corrections that are due to the vacuum polarization 
of the probe ({\it i.e.} highly virtual) photon in the definition of the measured form factors, 
the remaining QED radiative corrections to these form factors 
would be smaller than~0.5\%. 
The vacuum polarization and all other corrections are of opposite signs and partially cancel each other.}. 
Finally, we should emphasize that in order to account for the QED radiative corrections 
in a consistent manner we should have had these corrections implemented 
in the MC event generator that we use to measure the detection efficiency. 
We did not use such an event generator in our analysis. 

The efficiencies of the event selection criteria employed 
in our analysis are not flat over the studied $\qsq$ region. 
Most systematic errors for these efficiencies 
are quoted for a region of low $\qsq$ ({\it i.e.} less than $3~\gevt$) 
where the efficiencies are smaller and the systematic uncertainties are larger 
than at high $\qsq$ (above $3~\gevt$). 
These estimates are conservative in the high $\qsq$ region 
where a small fraction of signal events has been detected. 

In the analyses of $\gaga$ final states 
the systematic errors contribute a~5\% uncertainty 
to the measured cross sections. 
As we described above, 
this uncertainty includes a contribution of~$\sim3\%$ 
that comes from different sources 
for energy- and track-tagged event samples. 

\section{Analyses of Single-Tagged $6\gamma$ and $10\gamma$ Final States}

The following subsections mainly 
describe the differences among 
the analyses of $6\gamma$ and $10\gamma$ final states 
and the previously described analyses of $2\gamma$ final states, 
since they share many common features. 

\subsection{Trigger and Analysis Procedure}

In addition to the trigger utilized for $\gamma\gamma$ final states, 
$6\gamma$ and $10\gamma$ single-tagged events have been collected 
with a modified energy-based L1 trigger, which is fulfilled 
when the high energy shower associated with the 
tag candidate is found in the endcap calorimeter and 
two well-separated clusters, each of detected energy above 100 MeV, 
are identified in the barrel calorimeter. 
This additional trigger option is especially important for 
$3\pi^{0}$$\ra$$6\gamma$ and 
$5\pi^{0}$$\ra$$10\gamma$ final states 
because few of the photons resulting from 
the $\pi^{0}$ decays have sufficient energy  
to satisfy the high-energy trigger threshold of about $500$ MeV. 

Each event candidate should contain a tag (in the endcap part of the calorimeter), 
six, seven, ten or eleven photon candidates, and 
no charged tracks except the tag's track, if reconstructed. 
The efficiencies of these basic selection requirements 
are~30\%,~31\% and~12\% for the $\etaz \ra 6\gamma$, $\etap \ra 6\gamma$, and $\etap \ra 10\gamma$ 
analyses, respectively,  with the reconstruction efficiency of about~$80\%$ per photon 
being the dominant source. 

To reduce the systematic uncertainty in the trigger efficiency we select events 
with the most energetic photon candidate detected 
in the barrel calorimeter at polar angles above~$37^{\circ}$. 
The trigger efficiency for these events is larger than~90\%. 
We apply the same missing energy-momentum cut of~2.3~GeV as in the $\gaga$ analyses. 
We require acoplanarity less than~$30^{\circ}$ 
and do not apply a decay angle cut 
because there is no need to suppress the small background 
due to radiative Bhabha events. 

Only events that contain at least one combination of the required 
number of $\piz \ra \gaga$ and $\etaz \ra \gaga$ candidates 
are accepted for further analysis. 
To give an example, we consider the decay chain $\eta \ra 3\piz \ra 6\gamma$. 
Among six or seven photon candidates, there must be at least one 
set of three $\piz$ candidates, where each $\piz$ candidate 
is identified within \mbox{[-9.0, 3.5]$\sigma$} of the nominal $\piz$ mass. 
The mass resolution $\sigma$ 
has been measured as a function of energy and polar angle from data, 
with a typical value between 6 and 8 MeV/c$^2$. 
If there is more than one way to form three $\piz$ candidates, 
we use the best combination, {\it i.e.} the one which has the smallest $\chi^2$, where 
\begin{equation} 
%\[
\label{EQ:2001}
\chi^2 = \sum_{i=1}^{3} {\frac{(M_{\gamma\gamma}^{i}-
M_{\piz})^2}{\sigma^2_i}}. 
%\]
\end{equation} 
We follow the same procedure for $6\gamma$ and $10\gamma$ final states 
in which we search for the best $\piz\piz\etaz$ and $5\piz$ combinations, respectively. 
To obtain a better estimate of the parent particle four-momentum 
we perform a kinematic fit for each $\gaga$-decay 
candidate from the best combination. 
For events in which we find an additional energy cluster 
that has not been used to form any of the $\piz$ or $\etaz$ candidates, 
we require that the energy of this cluster be less than~200 MeV. 
In contrast to the $\gaga$ analysis, 
this energy cluster is not necessarily the least energetic one. 

Events that are accepted for further analysis must 
have constrained values of the tag scattering angle 
larger than~$15^{\circ}$. 
In addition, the detected fraction of the tag energy must be at least~50\%. 
To estimate the constrained values of the tag energy and scattering angle 
we employ energy and momentum conservation laws in which 
we use the four-momenta of the reconstructed 
$\piz\ra\gaga$ and $\etaz\ra\gaga$ candidates obtained from the kinematic fits. 

\begin{figure}
\centerline{
	\psfig{figure=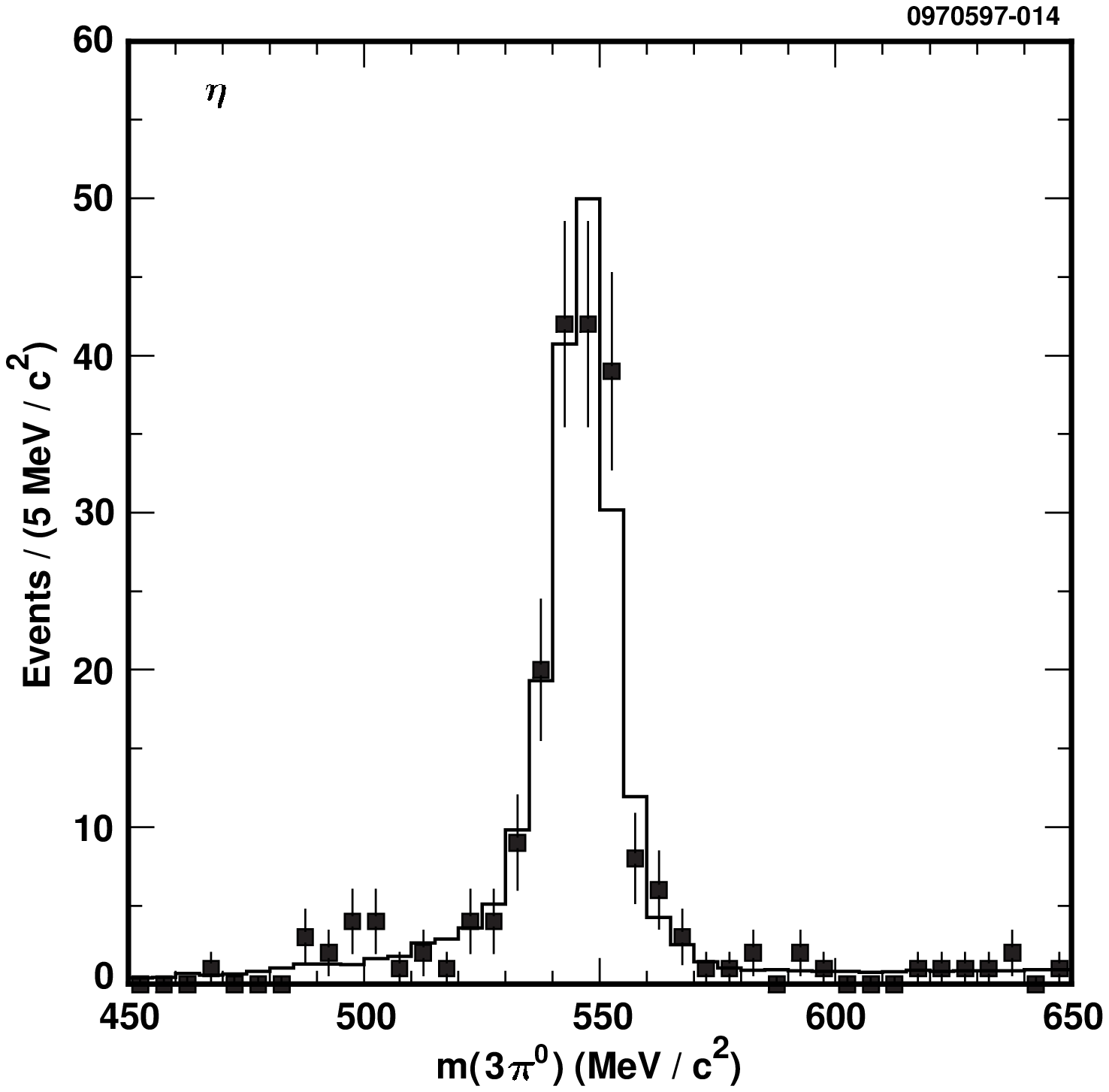,height=3.00in}
           }
\caption{
Fit (solid line) to the $3\piz$ invariant mass distribution 
observed in data (points with error bars) in the $\etaz \ra 6\gamma$ analysis. 
The signal line shape is obtained from the MC simulation, 
the remaining random background is approximated by 
a first-order polynomial. 
}
\label{fig:fig_06}
\end{figure}

\begin{figure}
\centerline{
	\psfig{figure=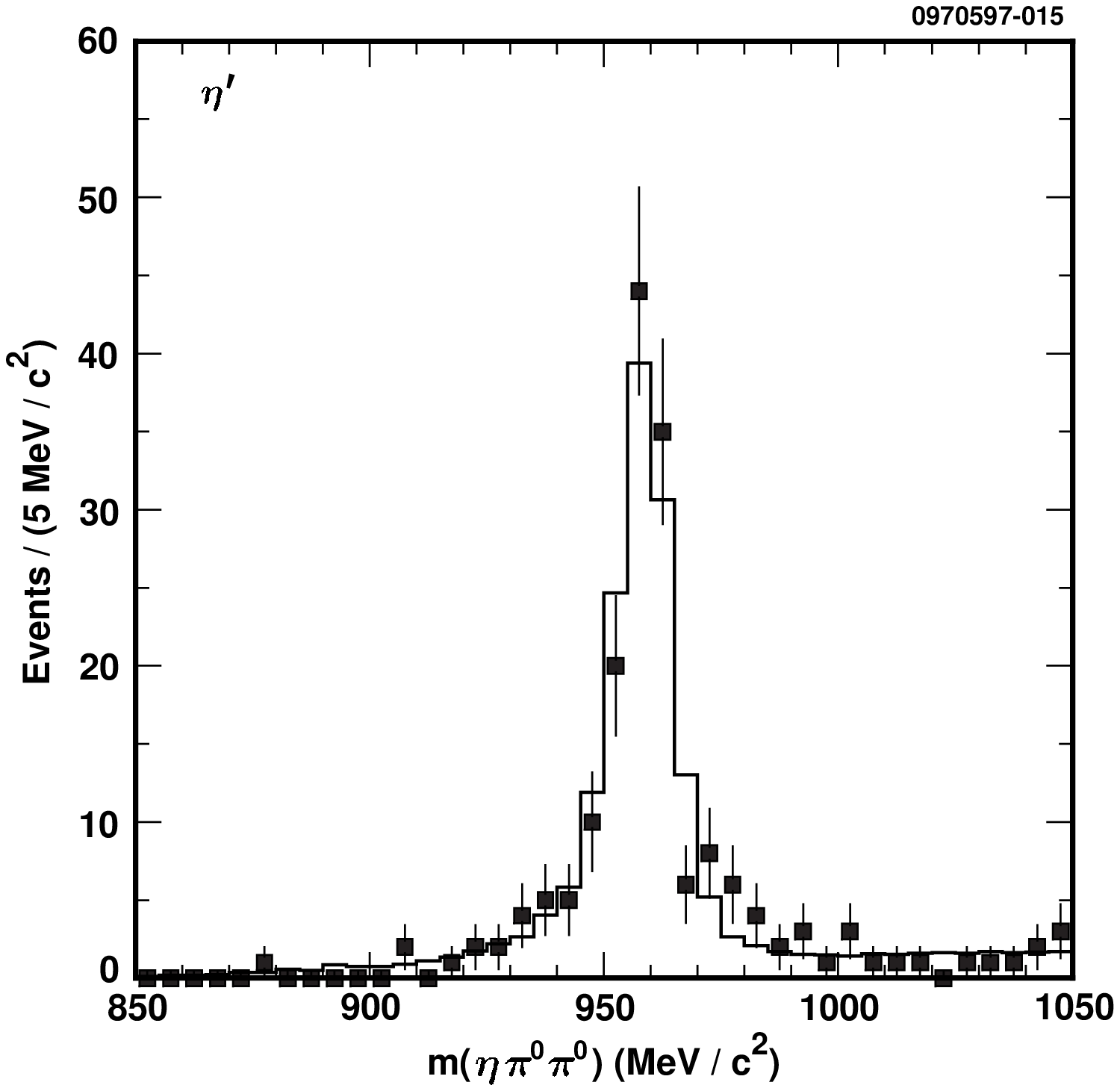,height=3.00in}
           }
\caption{
Fit (solid line) to the \mbox{$\piz\piz\etaz$ $\ra$ $6\gamma$} invariant mass distribution 
observed in data (points with error bars) in the $\etap \ra 6\gamma$ analysis. 
The signal line shape is obtained from the MC simulation, 
the remaining random background is approximated by 
a first-order polynomial. 
}
\label{fig:fig_07}
\end{figure}

\begin{figure}
\centerline{
	\psfig{figure=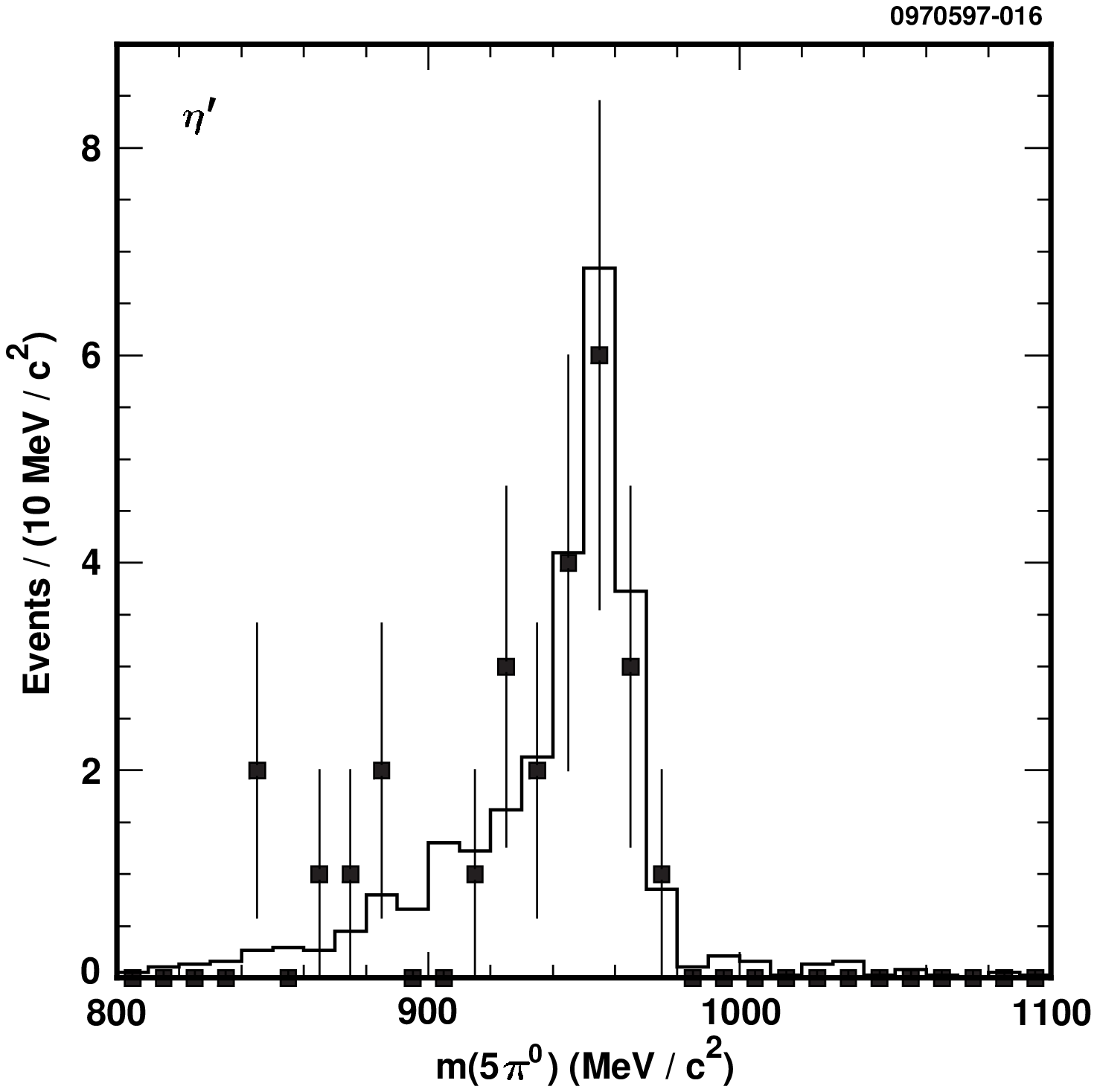,height=3.00in}
           }
\caption{
Fit (solid line) to the $5\piz$ invariant mass distribution 
observed in data (points with error bars) in the $\etap \ra 10\gamma$ analysis. 
The signal line shape is obtained from the MC simulation, 
the remaining random background is approximated by 
a first-order polynomial. 
}
\label{fig:fig_08}
\end{figure}

In Figs.~\ref{fig:fig_06}--\ref{fig:fig_08} 
we show the invariant mass distributions 
for data events that pass all selection criteria 
for the $\etaz$ and $\etap$ candidates 
and have values of $\qsq$ between~1.5 and~9 $\gevt$. 
The points with error bars in these figures 
represent event yields in data. 
The solid line in each figure shows the result of the binned likelihood fit 
to data with the signal line shape obtained from the MC simulation 
and a first-order polynomial chosen to approximate the remaining random background. 

\subsection{Background Estimates and Systematics}

To estimate the feed-down background, 
we have studied the distribution of extra energy 
when the cut on this quantity has been removed. 
We conclude that out of 187 
event candidates for the decay \mbox{$\etaz \ra 3\piz$} 
in data, 7 events are due to feed-down 
from the decay chain $\etap \ra \piz\piz\etaz \ra 5\piz$. 
To subtract this feed-down background, we use 
the extra energy spectra measured from data and signal MC simulation. 
We do not observe a feed-down in the $\etap$ analyses. 
We estimate the beam-gas and $e^+e^-$ annihilation backgrounds to be 
less than~1\% of the signal in each analysis. 

In the analyses of $6\gamma$ and $10\gamma$ final states 
we include a~1\% error to the systematics due to the uncertainty in the efficiency 
of the barrel energy-based L1 trigger. 
To estimate this uncertainty we have studied 
the efficiency of a low-energy trigger threshold 
for signal data and MC events which have been inclusively 
triggered with a high-energy trigger threshold. 
All other systematic uncertainties have been discussed in Section~III.D. 

In the analyses of $6\gamma$ and $10\gamma$ final states 
the overall systematic uncertainties in the measured cross sections 
are~7\% and~11\%, respectively. 

\section{Analyses of Single-Tagged Final States with Charged Pions}

In this section we describe the analyses of final states that 
contain the tag, two or four charged pions, and at least one photon. 

\subsection{Trigger}

As we described in preceding sections, 
charged tracks can be reconstructed only 
in events which have been recorded with the track-based L0 trigger. 
This trigger is satisfied by two well-separated 
TF hits, or one TF hit and a VD track. 
The L0 triggers are not correlated with the L1 triggers; 
when any of the L0 triggers is satisfied, 
all L1 triggers are examined~\cite{CLEO-II:trigger}. 

In addition to the energy-based L1 trigger described 
previously,  there are several track-based L1 triggers 
which are efficient for events with charged particles: 
\begin{itemize}
\item
The L1 ``electron'' 
trigger is satisfied by a high-threshold bit 
in the barrel calorimeter and a charged 
track penetrating more than halfway through the volume of the 
main drift chamber. 

\item 
The L1 ``two-track'' trigger is efficient for events 
with two or more low transverse-momentum 
charged particles; it 
requires at least 
two hits in either region of the TF system, 
two well-separated low-threshold clusters 
in the barrel calorimeter, and 
two charged tracks, each of transverse momentum 
above~90 MeV/c. 
The L2 trigger is fulfilled when at least one charged track 
of transverse momentum larger than~340 MeV/c is identified. 

\item

The ``hadronic'' triggers are designed for multi-particle 
final states from $e^{+}e^{-}$ annihilation, but have significant 
efficiency for this analysis as well. 
These have a variety of possible criteria involving the 
drift chambers, TF, and low-threshold bits of the 
calorimeter. In general, at least three tracks are required. 
\end{itemize}

Associated with these track-based L1 triggers, 
earlier data sets 
had a L2 requirement of a VD hit pattern 
consistent with a charged track of transverse momentum larger than~125 MeV/c. 
The LVL3 filter does not reject events that are 
collected by the track-based L1 triggers. 

\subsection{Analysis Procedure}

Each event candidate must contain the tag, 
an exact number of charged tracks 
(excluding the tag's track, if reconstructed), 
and at least as many photon candidates as 
are needed for full reconstruction of a studied decay chain. 
All tracks except for the tag's track 
are assumed to be due to charged pions. 
The net charge of the reconstructed pions must be zero. 
Photon candidates include all barrel (endcap) calorimeter clusters 
of energies larger than 30 (50) MeV except for those 
that are closest to the intersection points 
of charged tracks with the calorimeter. 
The efficiencies of the basic requirements described above 
are defined by the charged pion and photon reconstruction 
efficiencies, each about~$80\%$ per particle. 

To select events that trigger with high efficiency 
and small systematic uncertainty, 
we impose several event-quality criteria. 
Namely, we require that at least one charged 
track of transverse momentum larger than~250~MeV/c be 
detected. 
In addition, 
the charged track of largest transverse momentum 
and either the tag or the most energetic photon candidate 
must be detected in the barrel calorimeter at polar angles above~$37^{\circ}$. 
Finally, we reject events which contain charged tracks 
of momenta less than~80~MeV/c 
because for these tracks 
the systematic uncertainty in the track reconstruction efficiency is large. 

Given that $\rhoz\gamma$ events are primarily 
recorded with the energy-based L1 trigger, 
tighter event selection criteria are imposed 
in this analysis. 
We select events which have at least one charged 
track of transverse momentum above~450~MeV/c. 
The most energetic photon candidate 
must have energy, $E_{\gamma}$, larger than~130~MeV. 
We assume this photon candidate to be due to the 
signal process $\etap \ra \rhoz\gamma$. 
To suppress random background we select events 
with the reconstructed $\pip\pim$ mass 
between~550 and~800~MeV/c$^2$. 
This is referred to as the $\rhoz$-mass cut. 

In the analyses of the final states that contain 
the decays $\piz \ra \gaga$ and $\etaz \ra \gaga$, 
events must contain at least one combination 
of the exact number of the candidates for these decays 
as required for full reconstruction of the studied decay chain. 
The energy clusters that enter the best combination 
are assumed to be signal photons. 
The total energy collected in the calorimeter clusters other 
than the signal photon candidates and the energy clusters matched 
to the projections of the charged tracks must be less than 500 MeV. 
These extra energy clusters are mostly due to the interactions of the 
charged pions with the materials of the detector. 
No requirement is made on the number of such clusters. 

We use the momenta of the charged tracks and signal photons 
(after kinematic fits, where applicable) 
and employ energy-momentum conservation  
to estimate the tag energy and scattering angle. 
We select events in which the detected fraction of the tag energy 
is at least~50\% 
and a scattering angle is larger than~$15^{\circ}$ where both 
parameters are estimated using energy-momentum conservation. 

In the \mbox{$\etaz \ra \pip\pim\piz$} analysis we need to suppress 
a large feed-down from the decay chain $\etap \ra \piz\piz\etaz \ra \pip\pim3\piz$. 
To suppress this feed-down, 
we require the difference between the measured and constrained values 
of the tag scattering angle be less than~$2^{\circ}$. 
The feed-down suppression power of the combination of this and the extra energy cuts 
is a factor of~23, while the efficiency loss is less than~3\%. 

We employ the particle identification capabilities of our 
apparatus to reduce the large random background observed 
in the \mbox{$\etap \ra \rhoz\gamma$} analysis. 
This random background is primarily due to the process \mbox{$e^+e^- \ra e^+e^-e^+e^-$} 
accompanied by bremsstrahlung radiation, split-off showers or 
beam-related energy clusters. 
To suppress random background 
we utilize the fact that 
specific ionization energy losses, $dE/dx$, 
are larger for electrons than for charged pions. 
This information is used in the requirement on $P_{\chi^2}$, 
the upper tail probability of the {$\chi^2$} distribution 
of the $dE/dx$ measurements for charged pion candidates~\cite{UTP}. 
In the ideal case ({\it i.e.} if the $dE/dx$ distribution were Gaussian) 
the correct choice of the particle-identification hypothesis 
would produce a uniform $P_{\chi^2}$ distribution, 
while 
events with an incorrect particle-identification hypothesis 
tend to congregate near zero. 
We calculate $P_{\chi^2 }$ for the tracks assuming them to be due to charged pions. 
To suppress unwanted background events, 
$P_{\chi^2}$ is required to be larger than~0.005. 
The efficiency of the $P_{\chi^2}$ cut is not~99.5\% but~98\% 
because a small fraction of the signal events (in both data and simulation) 
does not have $dE/dx$ information and the $dE/dx$ distribution 
has non-Gaussian tails. 
The same cut on $P_{\chi^2 }$ is applied in all analyses 
with charged pions. 

\begin{figure}
\centerline{
	\psfig{figure=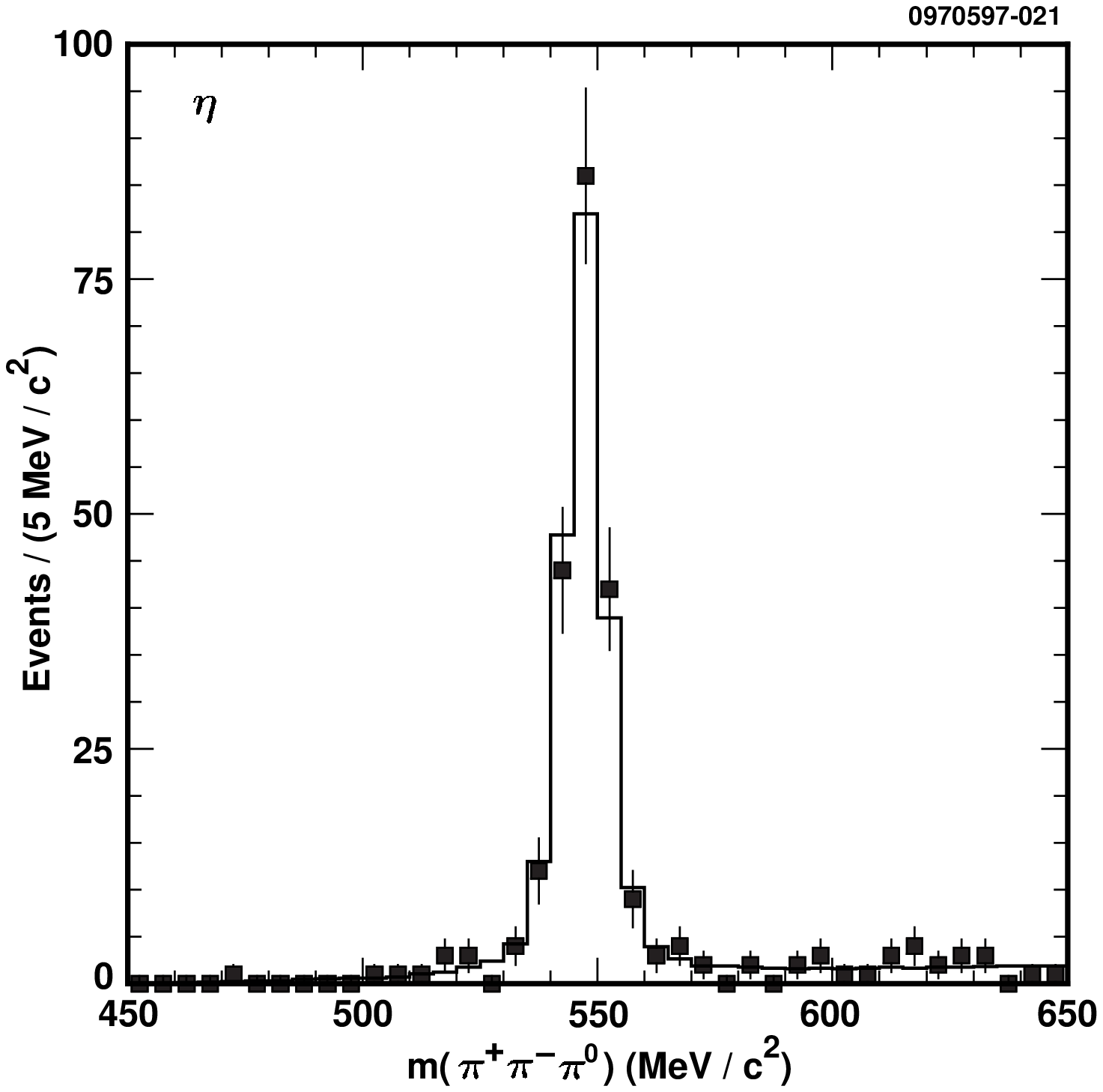,height=3.00in}
           }
\caption{
Fit (solid line) to the $\pip\pim\piz$ invariant mass distribution 
observed in data (points with error bars) in the $\etaz \ra \pip\pim2\gamma$ analysis. 
The signal line shape is obtained from the MC simulation, 
the remaining random background is approximated by 
a first-order polynomial. 
}
\label{fig:fig_09}
\end{figure}

\begin{figure}
\centerline{
	\psfig{figure=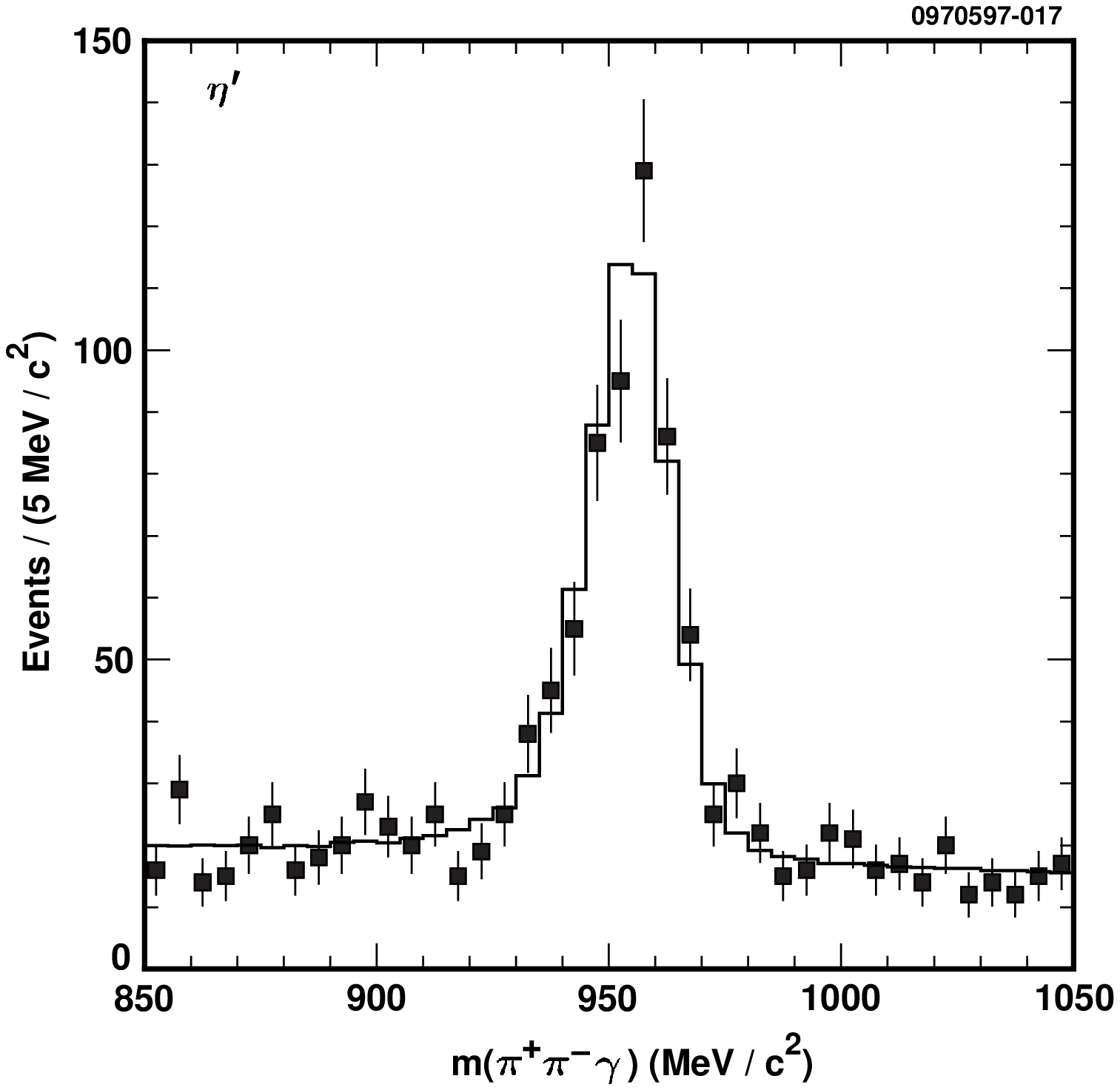,height=3.00in}
           }
\caption{
Fit (solid line) to the $\pip\pim\gamma$ invariant mass distribution 
observed in data (points with error bars) in the $\etap \ra \pip\pim\gamma$ analysis. 
The signal line shape is obtained from the MC simulation, 
the remaining random background is approximated by 
a first-order polynomial. 
}
\label{fig:fig_10}
\end{figure}

\begin{figure}
\centerline{
	\psfig{figure=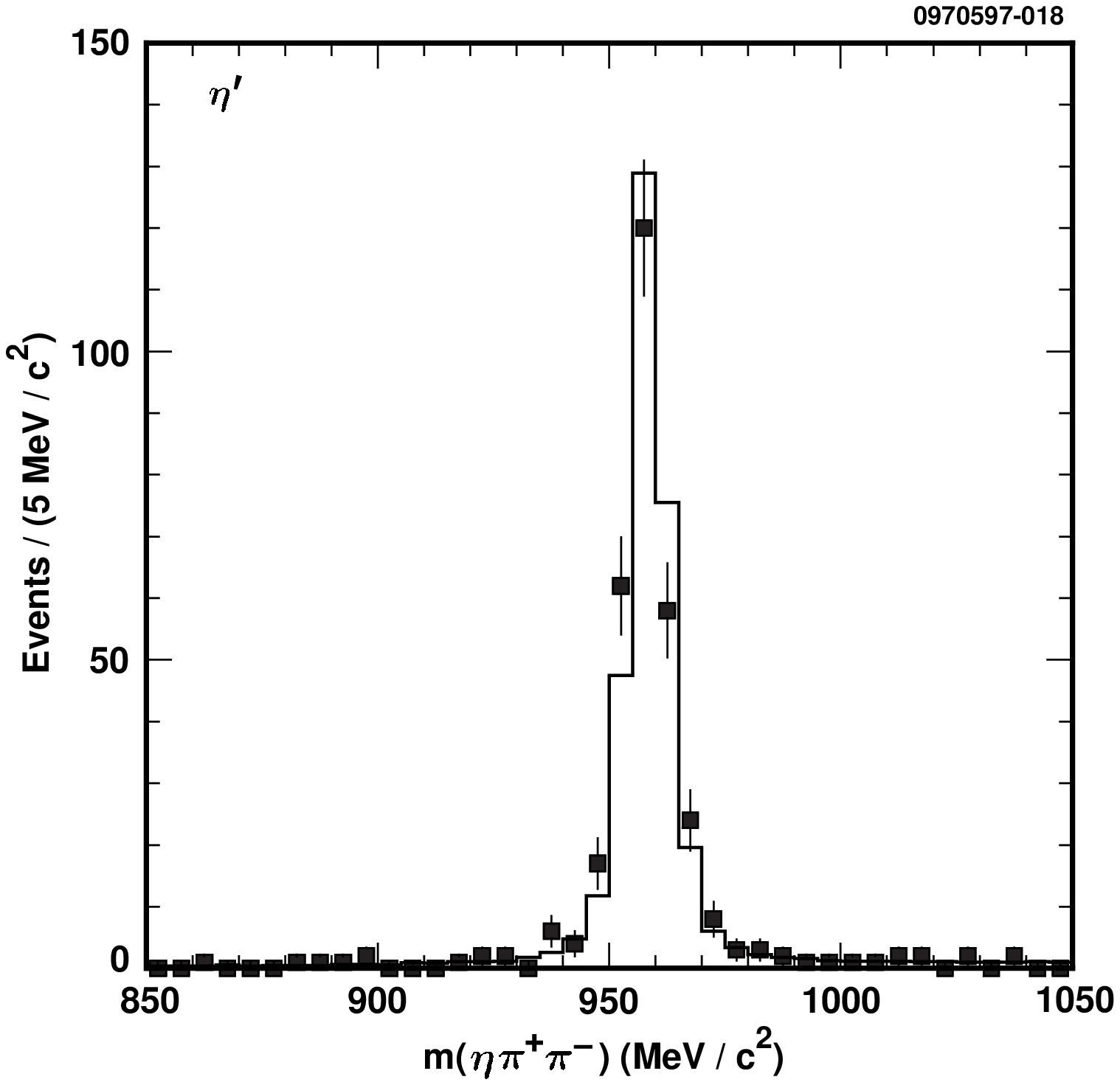,height=3.00in}
           }
\caption{
Fit (solid line) to the \mbox{$\pip\pim\etaz$ $\ra$ $\pip\pim2\gamma$} invariant mass distribution 
observed in data (points with error bars) in the $\etap \ra \pip\pim2\gamma$ analysis. 
The signal line shape is obtained from the MC simulation, 
the remaining random background is approximated by 
a first-order polynomial. 
}
\label{fig:fig_11}
\end{figure}

\begin{figure}
\centerline{
	\psfig{figure=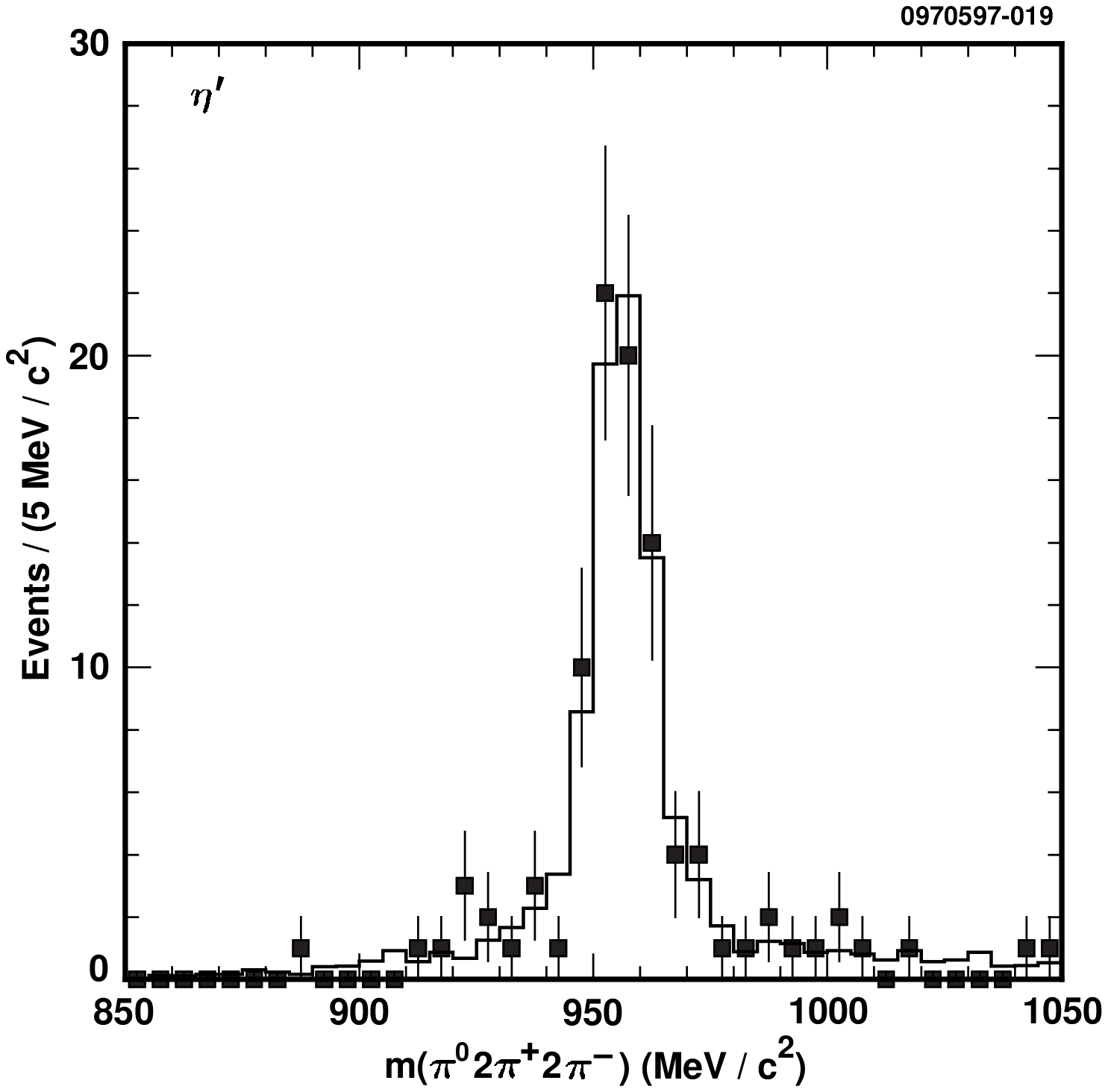,height=3.00in}
           }
\caption{
Fit (solid line) to the \mbox{$\pip\pim\etaz$ $\ra$ $2\pip2\pim2\gamma$} invariant mass distribution 
observed in data (points with error bars) in the $\etap \ra 2\pip2\pim2\gamma$ analysis. 
The signal line shape is obtained from the MC simulation, 
the remaining random background is approximated by 
a first-order polynomial. 
}
\label{fig:fig_12}
\end{figure}

\begin{figure}
\centerline{
	\psfig{figure=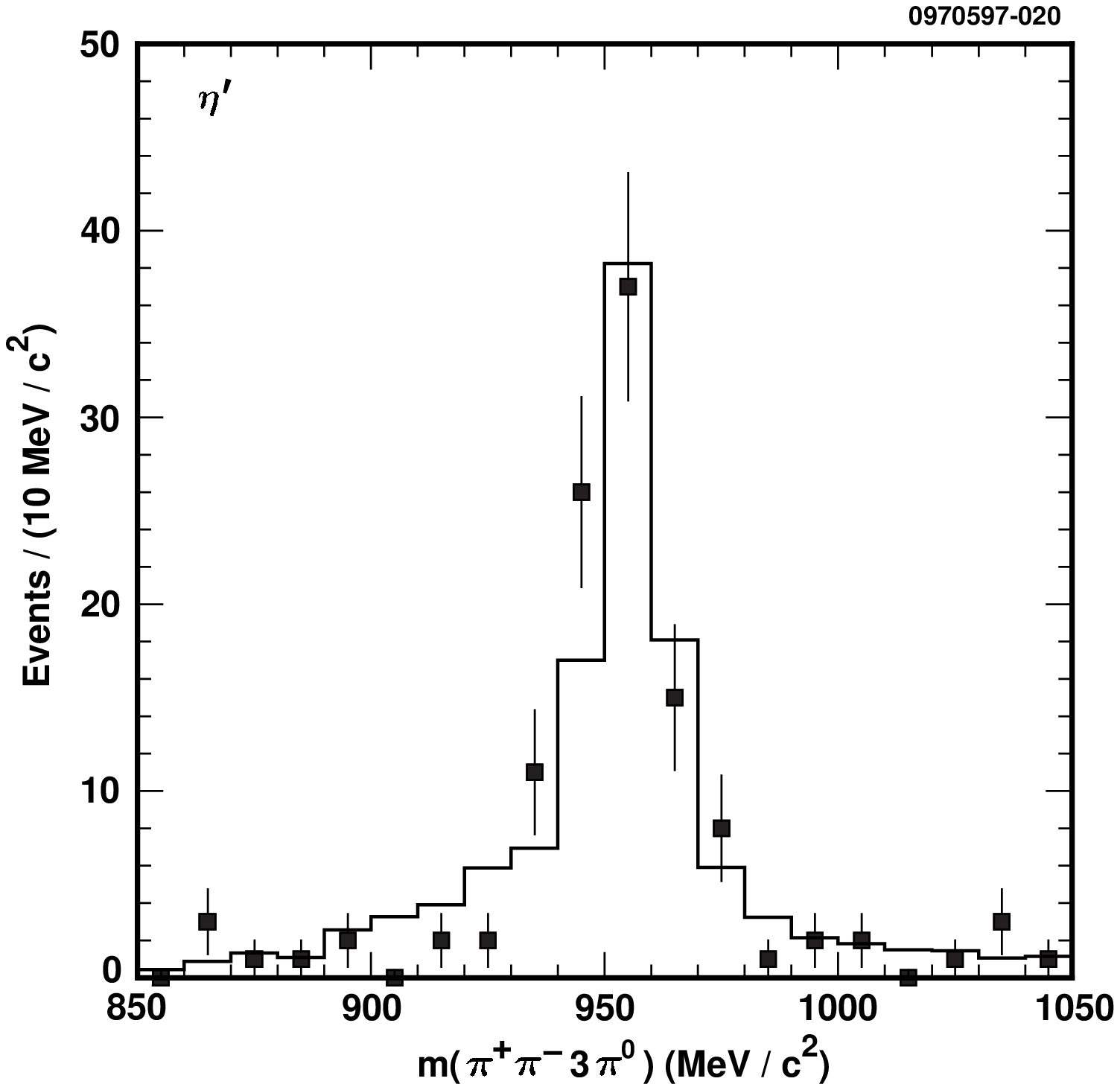,height=3.00in}
           }
\caption{
Fit (solid line) to the $\pip\pim3\piz$  invariant mass distribution 
observed in data (points with error bars) in the $\etap \ra \pip\pim6\gamma$ analysis. 
The signal line shape is obtained from the MC simulation, 
the remaining random background is approximated by 
a first-order polynomial. 
}
\label{fig:fig_13}
\end{figure}

In Figs.~\ref{fig:fig_09}--\ref{fig:fig_13} 
we show the invariant mass distributions 
for data events that pass all selection criteria 
for the $\etaz$ ($\etap$) candidates 
and have values of $\qsq$ between 1.5 and 20 (30) $\gevt$. 
The points with error bars in these figures 
represent event yields in data. 
The solid line in each figure shows the result of the binned likelihood fit 
to data with the signal line shape obtained from the MC simulation 
and a linear approximation of the remaining random background. 
The remaining random background observed in the analysis of the $\pip\pim\gamma$ final state 
is due to the process $e^+e^- \ra e^+e^-\mu^+\mu^-$ 
accompanied by noise and split-off energy clusters. 

\subsection{Background Estimates and Systematics}

To estimate the feed-down background, 
we have analyzed the distributions of extra energy and 
the difference between the measured and constrained values 
of the tag scattering angle when the cuts on these quantities 
have been removed. 
We conclude that fewer than~2 events 
in the $\etaz \ra \pip\pim\piz$ analysis are due to feed-down 
from the decay chain $\etap \ra \piz\piz\etaz \ra \pip\pim3\piz$. 
We have not identified any feed-down background in the $\etap$ analyses. 
We estimate that the background contribution from beam-gas 
interactions and $e^+e^-$ annihilation processes 
is less than~1\% of the signal in all analyses. 
This gives an estimate of the relevant systematic uncertainty. 

To estimate the systematic uncertainty in the efficiency of the L0 trigger, 
we select signal events that are triggered by the TF-based L0 trigger 
and measure the VD efficiency per event. 
Using a similar method we measure the efficiency of the TF-based L0 trigger 
for events that are triggered by the track-based L0 trigger.  
We estimate the uncertainty in the L0 trigger efficiency to be~1\%, 
which is the typical deviation between either of these efficiencies 
measured from data and simulation. 
Note that the efficiency of the L0 trigger is higher than~98\% 
for events which satisfy the basic selection requirements. 

We have measured the efficiency of the $P_{\chi^2 }$ cut 
in a nearly background-free environment using 
fitted mass distributions for 
signal events in data and MC simulation 
for the decay chain $\etap \ra \pip\pim\etaz \ra \pip\pim2\gamma$. 
We have found this efficiency to be~98\% (as discussed in the previous subsection) 
and use the~2\% statistical error of this measurement 
as an estimate of the systematic uncertainty. 

To estimate the uncertainty in the efficiency of the $E_{\gamma}$ cut 
in the $\etap \ra \rhoz\gamma$ analysis, 
we have measured this efficiency assuming that the energies 
of the reconstructed photons in the simulation 
are systematically shifted by~2\% of their nominal values. 
We have observed a relative change of~1\% in the 
efficiency of the $E_{\gamma}$ cut and 
this gives an 
estimate of its systematic uncertainty. 

We estimate the uncertainty in the track reconstruction 
efficiency to be~2\% per charged pion. 
It is determined from a global fit of the measured ratios 
of the $\etaz$ and $\etap$ branching fractions 
to their average values~\cite{PDG:96}. 

\begin{figure}
\centerline{
	\psfig{figure=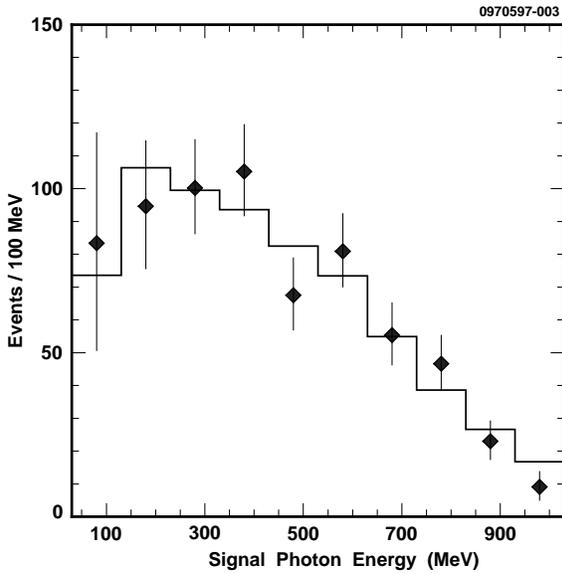,height=3.00in}
           }
\caption{
Distribution of signal photon energy 
in the $\etap \ra \rhoz\gamma$ analysis 
in data (points with error bars) 
and the MC simulation (histogram). 
The prediction of the MC simulation 
is normalized to the number of data events. 
}
\label{fig:fig_14}
\end{figure}

\begin{figure}
\centerline{
	\psfig{figure=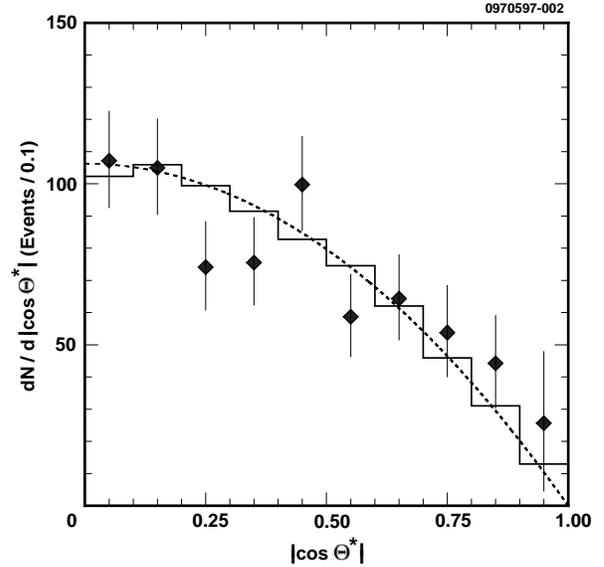,height=3.00in}
           }
\caption{
Distribution of $|\cos{\theta^*}|$ 
in the $\etap \ra \rhoz\gamma$ analysis 
in data (points with error bars) 
and the MC simulation (histogram). 
The dotted line shows the $\sin^2{\theta^*}$ curve. 
The prediction of the MC simulation and $\sin^2{\theta^*}$ curve 
are normalized to the number of data events. 
}
\label{fig:fig_15}
\end{figure}

The uncertainty in the efficiency of the $\rhoz$-mass cut is 
negligible because, except for the $\rhoz$-line shape, the matrix element 
for the decay chain \mbox{$\etap \ra \rhoz\gamma \ra \pip\pim\gamma$} 
is determined by QED and kinematics. 
To confirm this statement we remove 
the $\rhoz$-mass and $E_{\gamma}$ cuts 
and compare the distributions of 
$E_{\gamma}$ 
and 
$|\cos{\theta^*}|$ 
measured from signal data and MC simulation, where $E_{\gamma}$ is the 
signal photon energy in the lab frame\footnote{
These figures of merit for the analysis of the decay chain 
$\etap \ra \rhoz\gamma \ra \pip\pim\gamma$ 
were proposed in~\cite{PLUTO:ff}. 
}. 
These distributions are shown in Figs.~\ref{fig:fig_14}~and~\ref{fig:fig_15}. 
We observe good agreement between the data and MC spectra of   
$E_{\gamma}$ 
and 
$|\cos{\theta^*}|$ 
and conclude 
that the approximations given by Eqns.~\ref{EQ:EPRG1}~and~\ref{EQ:EPRG2} 
describe the data well. 
We note that both figures show the observed spectra, {\it i.e.} no 
detection efficiency corrections have been applied to these distributions. 
The good agreement between the shape of the $|\cos{\theta^*}|$ distribution 
obtained from the simulation 
and 
$\sin^2{\theta^*}$ curve is due to the detection efficiency 
being practically flat over the full range of $|\cos{\theta^*}|$. 

All other systematic uncertainties have been discussed 
in Sections~III.D~and~IV.B. 
In the analyses of final states with charged pions 
the overall systematic uncertainty in the measured cross sections 
is between~7\% and~10\%, depending on the final state. 

\section{Unfolding Procedure for the Transition Form Factors}

To measure the products of the differential cross sections 
and branching fractions for each decay chain 
we use the following analysis procedure. 
Data events that pass all selection criteria are used 
to form the $\qsq$ distribution where the value of $\qsq$ for 
each event is estimated from energy-momentum conservation 
(and the polar angle of the tag's track when the track is reconstructed). 
Next we divide the event yields into $\qsq$ intervals. 
For each $\qsq$ interval 
we obtain the number of signal events in data 
from the fit to the invariant mass distribution. 
Then we estimate and subtract the feed-down background 
using the methods described in preceding sections. 
Finally we 
correct the background-subtracted number of signal events 
for the detection efficiency. 
The signal line shapes used in the fits 
and the detection efficiencies 
are determined 
from the detector simulation 
for each $\qsq$ interval. 

To extract the transition form factors we compare 
the measured and the predicted values of the cross sections. 
Namely, for each $\qsq$ interval, we measure the form factors 
$\frg^{data}(\tilde{Q}^2)$ from: 
\begin{equation}
%\[
|\frg^{data}(\tilde{Q}^2)|^2 = \frac{\sigma(data)}{\sigma(MC)} |\frg^{MC}(\tilde{Q}^2)|^2,
%\]
\end{equation} 
where 
$\frg^{MC}(\tilde{Q}^2)$ is the approximation for the $\qsq$-dependent part 
of the form factor in MC simulation, 
and $\sigma(data)$ and $\sigma(MC)$ are the cross sections for 
this $\qsq$ interval measured in data and predicted using 
numerical integration, respectively. 
The transition form factors are measured at $\tilde{Q}^2$ where 
the differential cross sections achieve their mean values 
according to the results of numerical integration. 
The numerical results have been obtained 
at an average center-of-mass energy of 10.56 GeV 
with the approximation for the form factor given by Eqn.~\ref{EQ:137}. 

\begin{figure}
\centerline{
	\psfig{figure=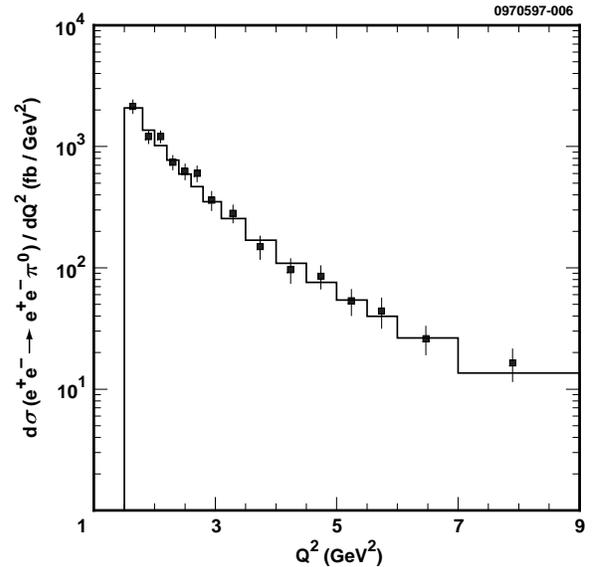,height=3.00in}
           }
\caption{
Measured (points with error bars) and numerically estimated (histogram) differential cross sections for $\piz$ production. 
}
\label{fig:fig_16}
\end{figure}

\begin{figure}
\centerline{
	\psfig{figure=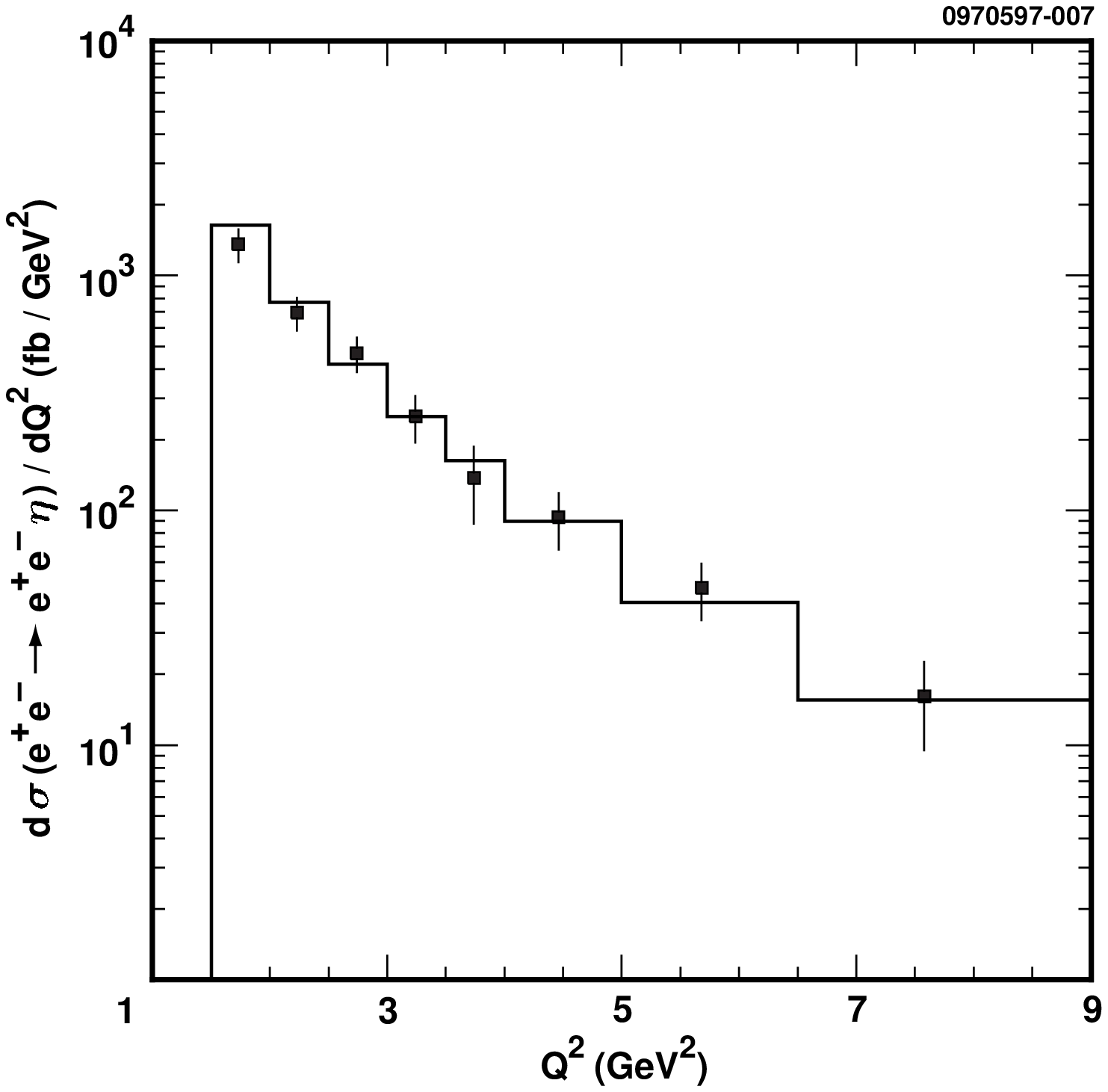,height=3.00in}
           }
\caption{
Measured (points with error bars) and numerically estimated (histogram) differential cross sections for $\etaz$ production 
in the $\etaz \ra \gaga$ analysis. 
}
\label{fig:fig_17}
\end{figure}

The $\qsq$ distributions measured from data and obtained numerically 
are shown in Figs.~\ref{fig:fig_16}~and~\ref{fig:fig_17} 
for the $\piz \ra \gaga$ and $\etaz \ra \gaga$ analyses, respectively. 
Only statistical errors are shown in these figures. 
To plot the results of numerical integration we use 
$\Gamma(\piz \ra \gaga) = 7.74 {\rm ~eV}$ 
and 
$\Gamma(\etaz \ra \gaga) = 463 {\rm ~eV}$ 
\cite{PDG:96}. 

We show our experimental results 
in Tables~I--X. These tables show 
the $\qsq$ intervals, 
event yields obtained from the fits, 
numbers of signal events after subtraction of the feed-down background, 
detection efficiencies, 
the $\tilde{Q}^2$ values, 
the products of the differential cross sections and relevant branching fractions, 
and 
the transition form factors, 
represented in the form $\tilde{Q}^2 |\frg(\tilde{Q}^2)|$. 
In Tables~I--X 
the first error is statistical 
and 
the second error (where given) is systematic. 

\section{Comparison of the Results with Theoretical Predictions}
 
In this section we compare the results for $\piz$ with theoretical predictions. 
For the transition form factors of $\etaz$ and $\etap$ 
we compare the results with the PQCD asymptotic prediction only 
because little is known in theory about the wave functions of these mesons. 
No predictions for the form factors of $\etaz$ and $\etap$ are available at this time 
except for the prediction of Kroll \etal ~\cite{KROLL:96} 
where these authors assumed that the shapes of the wave functions 
of all three pseudoscalar mesons are similar. 

\subsection{Results for $\piz$}

\begin{figure}
\centerline{
	\psfig{figure=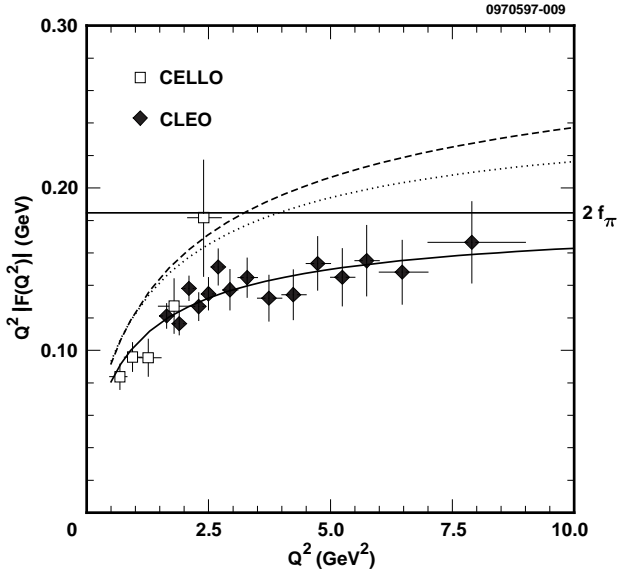,height=3.00in}
           }
\caption{Comparison of the results (points) for $\qsq|\fpizg(\qsq)|$ with 
the theoretical predictions made by Jakob \etal ~\protect\cite{KROLL:96} with 
the asymptotic wave function (solid curve) 
and 
the CZ wave function (dashed curve). 
The dotted curve shows the prediction made with 
the CZ wave function when its QCD evolution is taken into account. 
}
\label{fig:fig_18}
\end{figure}

\begin{figure}
\centerline{
	\psfig{figure=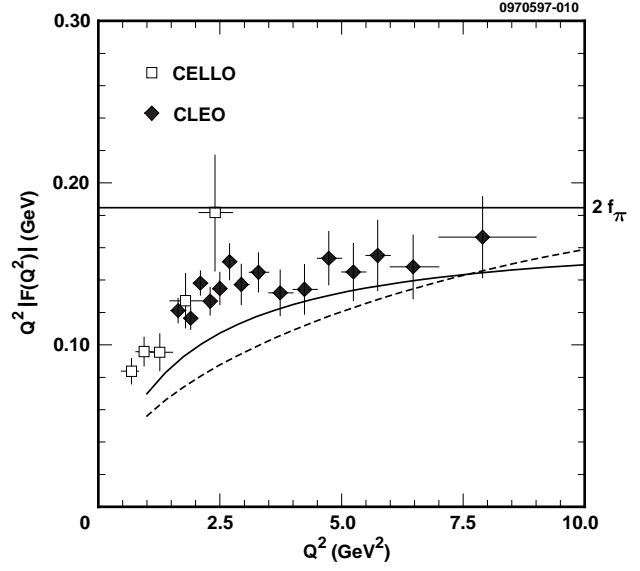,height=3.00in}
           }
\caption{Comparison of the results (points) for $\qsq|\fpizg(\qsq)|$ with 
the theoretical predictions made by Cao \etal ~\protect\cite{GUANG:96} with 
the asymptotic wave function (solid curve) 
and 
the CZ wave function (dashed curve). 
}
\label{fig:fig_19}
\end{figure}

\begin{figure}
\centerline{
	\psfig{figure=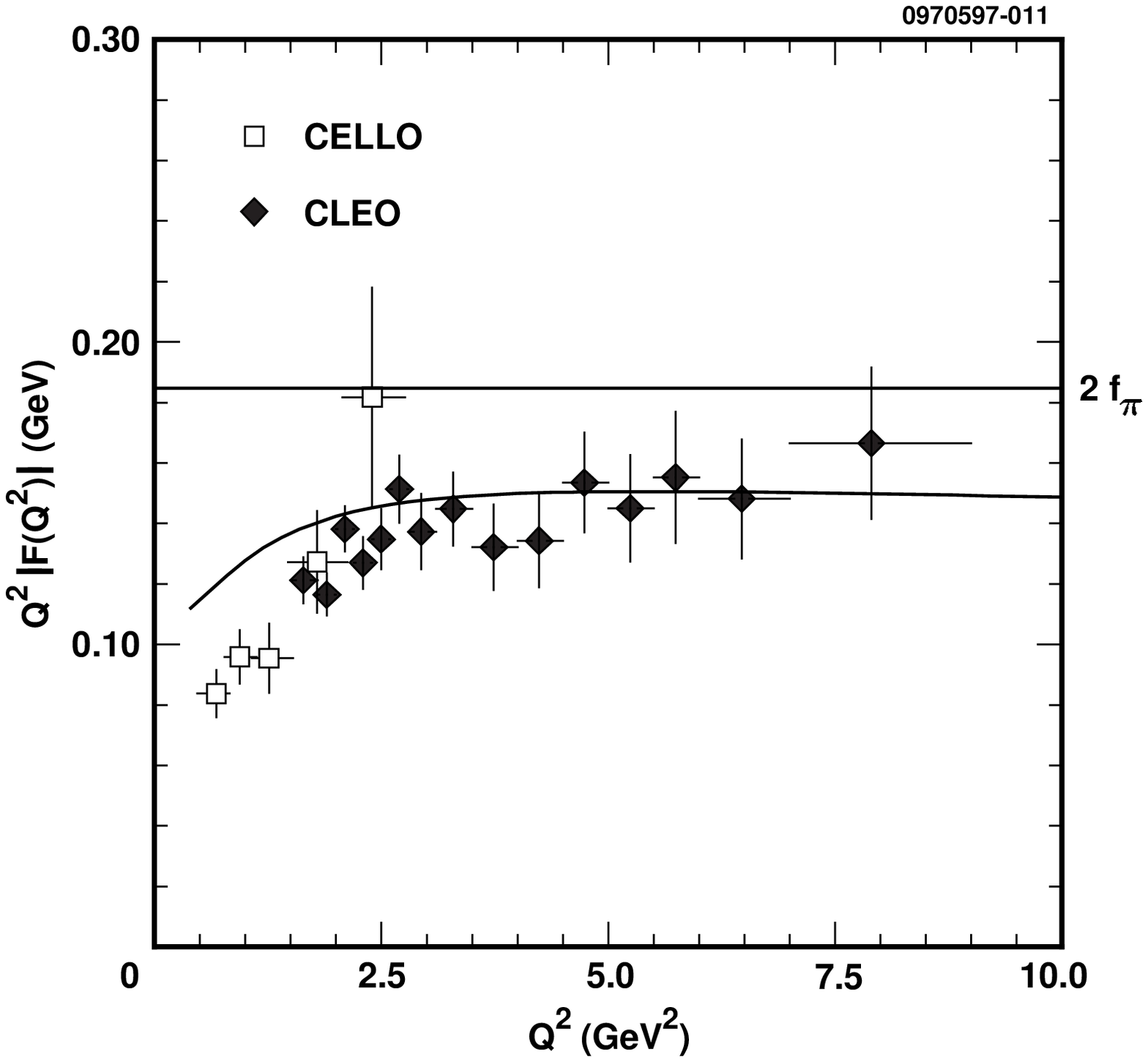,height=3.00in}
           }
\caption{Comparison of the results (points) for $\qsq|\fpizg(\qsq)|$ with the theoretical 
prediction (curve) made by Radyushkin \etal ~\protect\cite{RR:9603408}. 
}
\label{fig:fig_20}
\end{figure}

\begin{figure}
\centerline{
	\psfig{figure=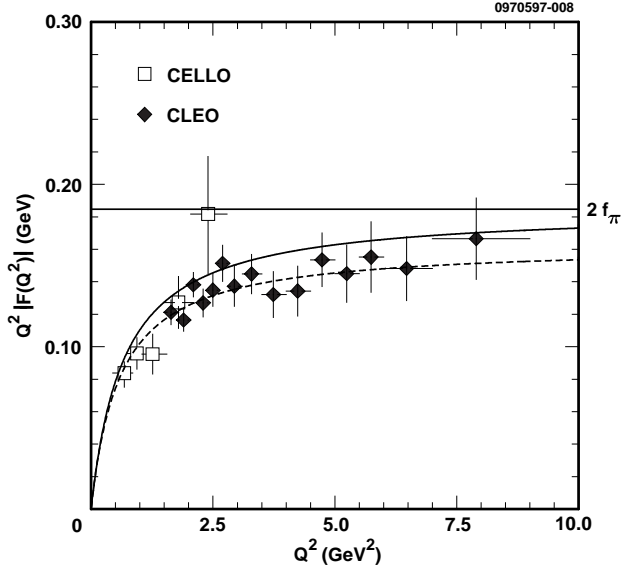,height=3.00in}
           }
\caption{
The interpolation given by Eqn.~\ref{EQ:45} (solid curve) and 
the pole-mass parameter fit (dashed curve) to our results (closed circles) for $|\fpizg(\qsq)|^2$ 
represented in the $\qsq |\fpizg(\qsq)|$ form. 
}
\label{fig:fig_21}
\end{figure}

In Figures~\ref{fig:fig_18}--\ref{fig:fig_21} 
we compare our results for $\qsq|\fpizg(\qsq)|$ with the theoretical predictions. 
Also shown in these figures are the results of the CELLO 
experiment~\cite{CELLO:ff} and the asymptotic prediction 
of PQCD given by Eqn.~\ref{EQ:41}. 
For both experimental results the error bars represent 
the statistical errors only. 
To plot the results of the theoretical predictions we use 
their published analytical forms. 
To estimate the value of $f_{\pi}$ we use Eqns.~\ref{EQ:37}~and~\ref{EQ:319} 
and the tabulated two-photon partial width of $\piz$~\cite{PDG:96}. 
This estimate of $f_{\pi}$ (92.3 MeV) agrees with its experimental value (92.4 MeV) 
which has been measured previously from charged pion decays\footnote{
For each meson ${\cal R}$, where ${\cal R}$ is $\piz$, $\etaz$ or $\etap$, 
our definition of the meson decay constant $f_{\cal R}$ 
differs by a factor of $1/\sqrt{2}$ from the one 
accepted by the Particle Data Group and given in~\cite{PDG:96}.}~\cite{PDG:96}. 

In Fig.~\ref{fig:fig_18} 
the results are compared with the predictions made by 
Jakob \etal ~\cite{KROLL:96}. 
These authors calculated the $\mgsg$ transition form factor 
by employing a PQCD-based technique and QCD radiative corrections~\cite{LS:92}. 
They used two estimates for the $\piz$ wave function: 
the asymptotic wave function 
and 
the CZ wave function. 
This theoretical prediction 
gives a much better agreement with our results 
when the asymptotic wave function is used. 
In terms of the PQCD-based approach this indicates that 
the wave function has already evolved to the asymptotic 
form at $\qsq$ as small as~1~$\gevt$. 
Notice that $\fpizg$ calculated with the CZ wave function 
changes when the QCD evolution of this wave function 
over the studied $\qsq$ range  
is taken into account according to~\cite{KROLL:WUB9619}. 
The transition form factor does not change when the asymptotic wave 
function is used because this wave function exhibits no 
QCD evolution to leading order in $\alpha_s$. 
However, 
in next-to-leading order in $\alpha_s$ any wave function, including the asymptotic, 
is subject to the QCD evolution~\cite{MULLER:95}. 
If this evolution is taken into account 
the prediction with the asymptotic wave function 
which has been derived to leading order in $\alpha_s$ 
would also change slightly~\cite{KROLL:WUB9619}. 

Cao \etal ~also made a prediction based on PQCD~\cite{GUANG:96}. 
These authors disagreed with the approximations made 
to simplify the form of HSA in~\cite{KROLL:96}. 
Their prediction includes transverse momentum corrections 
and is compared with our results in 
Fig.~\ref{fig:fig_19} 
for the asymptotic and CZ wave functions. 
The theoretical prediction of Cao \etal ~yields 
a smaller value of $\fpizg$ for $\qsq$ less than~8~$\gevt$ 
when the CZ wave function is used. 
This is a most intriguing result 
because the CZ wave function has been proposed 
to account for measured excesses in the rates for various processes, 
thus leading to larger values of the form factors and cross sections~\cite{CZ:84}. 

The prediction of Radyushkin \etal ~\cite{RR:9603408} based on the QCD sum-rules 
method~\cite{AV:sum_rules} is compared with the experimental results 
in Fig.~\ref{fig:fig_20}. 
This calculation describes the saturating behavior of our 
measurement, though it disagrees with 
the data at smaller $\qsq$. 
It should be noted that at low $\qsq$ the prediction is not expected 
to agree with the data: the QCD radiative corrections 
which would be larger at smaller $\qsq$ 
have not been included in this theoretical analysis. 
The discrepancy between the absolute values of the asymptotic 
limits of PQCD and of this prediction might be due 
to the uncertainties in the expectation values of 
the vacuum condensates that are known only with~30\% precision~\cite{AV:sum_rules}. 
However, according to the authors, the agreement can be achieved 
by means of complicated QCD-evolution analysis of the 
correlator functions used in this theoretical approach~\cite{RR:comm_97}. 

Finally, we derive the value of the pole-mass parameter $\Lambda_{\piz}$ 
which we use to represent our results in a simple phenomenological form. 
We fit our results for $|\fpizg(\qsq)|^2$ with a function 
given by Eqn.~\ref{EQ:137} and obtain the following result: 
\begin{equation}
%\[
\Lambda_{\piz} = 776 \pm 10 \pm 12 \pm 16 {\rm ~MeV}, 
%\]
\label{EQ:1008}
\end{equation} 
where the first error is statistical, the second error represents systematic 
uncertainties of our measurements, and the third error is due to 
the uncertainty in the value of $\Gamma(\piz \ra \gaga)$~\cite{PDG:96}. 
The result of the fit is shown in Fig.~\ref{fig:fig_21}. 
While we observe that a simple VMD-like approximation describes the data very well, 
we should note that it disagrees with the asymptotic prediction of PQCD. 
Also shown in Fig.~\ref{fig:fig_21} is the interpolation given by Eqn.~\ref{EQ:45}. 

\subsection{Results for $\etaz$}

\begin{figure}
\centerline{
	\psfig{figure=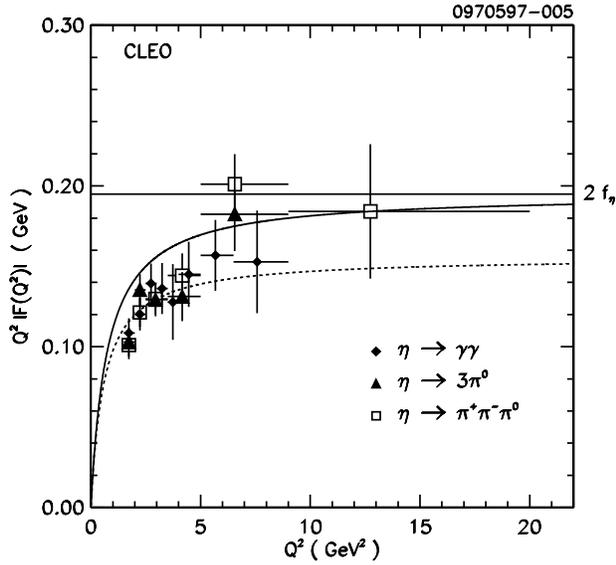,height=3.00in}
           }
\caption{
Results of the pole-mass parameter fit to our results (points) for $|\fetazg(\qsq)|^2$ 
represented in the $\qsq |\fetazg(\qsq)|$ form (dashed line).
The solid curve shows the interpolation given by Eqn.~\ref{EQ:45}. 
}
\label{fig:fig_22}
\end{figure}

We show the results of our measurements for $\qsq|\fetazg(\qsq)|$ 
in Fig.~\ref{fig:fig_22}. 
This figure also shows the asymptotic prediction of PQCD given by Eqn.~\ref{EQ:41} and 
the interpolation given by Eqn.~\ref{EQ:45}. 
To estimate the value of $f_{\etaz}$ (97.5 MeV) we use Eqns.~\ref{EQ:37}~and~\ref{EQ:319} 
and the tabulated two-photon partial width of $\etaz$~\cite{PDG:96}. 
We fit the $|\fetazg(\qsq)|^2$ distributions measured using each decay chain 
with the functional form given by Eqn.~\ref{EQ:137} and obtain 
the values of the pole-mass parameter $\Lambda_{\etaz}$ that are shown 
in Table~\ref{tab:final_res_pole}. 
In this table, for each measurement, the first error is statistical, 
the second error represents systematic uncertainties of our measurement, 
and the third error reflects 
the uncertainty in the two-photon partial width of $\etaz$. 
From a simultaneous fit to our three measurements for the production of $\etaz$ 
we obtain the following value of the pole-mass parameter: 
\begin{equation}
%\[
\label{EQ:98001}
\Lambda_{\etaz} = 774 \pm 11 \pm 16 \pm 22 {\rm ~MeV}. 
\end{equation}
The result of this fit is shown in Fig.~\ref{fig:fig_22}. 

We use the measured values of the parameters $\Lambda_{\piz}$ and $\Lambda_{\etaz}$ 
to compare the soft non-perturbative properties of $\piz$ and $\etaz$. 
This is a legitimate comparison 
because 
the chiral limit given by Eqn.~\ref{EQ:319} 
and 
the asymptotic prediction given by Eqn.~\ref{EQ:41} 
are expected to hold for both $\piz$ and $\etaz$. 
From the comparison between the measured values of $\Lambda_{\piz}$ and $\Lambda_{\etaz}$ 
we conclude that the $\qsq$ shapes of the \mbox{$\gamma^*\gamma$ $\ra$ meson} 
transition form factors of $\piz$ and $\etaz$ are nearly identical, 
which strongly indicates the similarity between the wave functions of these mesons. 

\subsection{Results for $\etap$}

\begin{figure}
\centerline{
	\psfig{figure=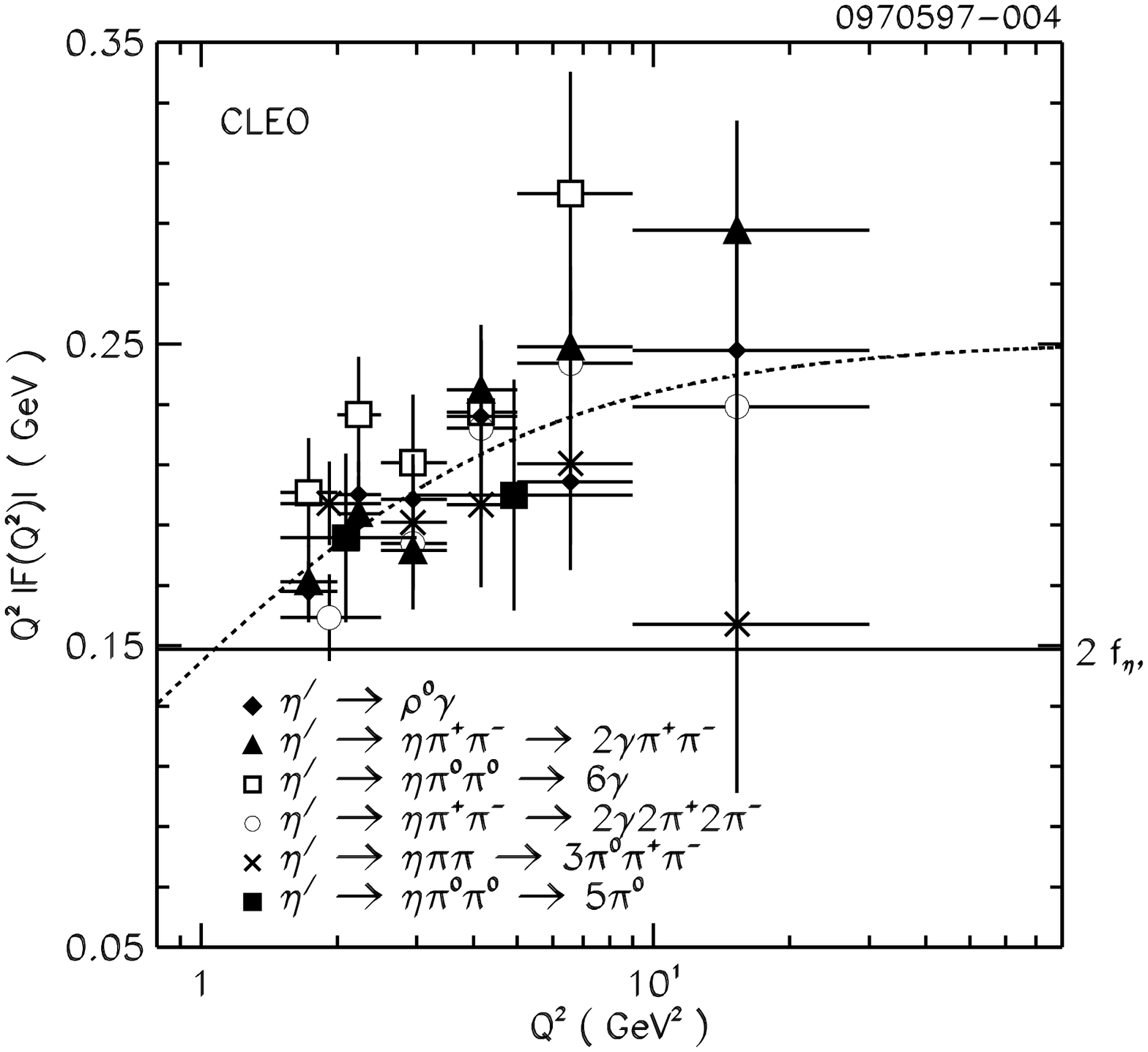,height=3.00in}
           }
\caption{
Results of the pole-mass parameter fit to our results (points) for $|\fetapg(\qsq)|^2$ 
represented in the $\qsq |\fetapg(\qsq)|$ form (dashed line).
}
\label{fig:fig_23}
\end{figure}

We show the results of our measurements for $\qsq|\fetapg(\qsq)|$ 
in Fig.~\ref{fig:fig_23}. 
This figure also shows 
what would be the PQCD asymptotic prediction given by Eqn.~\ref{EQ:41} 
for $\qsq|\fetapg(\qsq)|$ if the chiral limit given by Eqn.~\ref{EQ:319} 
held for $\etap$. 
To estimate the value of $f_{\etap}$ (74.4 MeV) we use Eqns.~\ref{EQ:37}~and~\ref{EQ:319} 
and the tabulated two-photon partial width of $\etap$ of 
$4.3 {\rm ~keV}$~\cite{PDG:96}. 

We fit the $|\fetapg(\qsq)|^2$ distributions measured using each decay chain 
with the functional form given by Eqn.~\ref{EQ:137} and obtain 
the values of the pole-mass parameter $\Lambda_{\etap}$ that are shown 
in Table~\ref{tab:final_res_pole}.
From a simultaneous fit to our six results for the production of $\etap$ 
we obtain the following value of the pole-mass parameter: 
\begin{equation}
%\[
\label{EQ:98002}
\Lambda_{\etap} = 859 \pm 9 \pm 18 \pm 20 {\rm ~MeV}.
\end{equation}
The result of this fit is shown 
in Fig.~\ref{fig:fig_23}. 

The results of our measurements for the production of $\etap$ 
demonstrate that 
if this particle were a $q\bar{q}$ bound state 
and 
the QCD chiral limit given by Eqn.~\ref{EQ:319} held for this meson, 
the $\qsq$-dependence of the transition form factor of $\etap$ 
and consequently its wave function would be significantly 
different from these non-perturbative properties of either~$\piz$~or~$\etaz$. 

\section{Conclusions}

We have measured the form factors associated with 
the electromagnetic transitions \mbox{$\gamma^*\gamma$ $\ra$ meson} 
in the regions of momentum transfer from 1.5 to 
9, 20, and 30 $\gevt$ 
for the $\piz$, $\etaz$, and $\etap$ mesons, respectively. 
These are the first measurements 
above 2.7 $\gevt$ for $\piz$ and above 7 $\gevt$ 
for $\etaz$ and $\etap$. 

Our measurement for $\piz$ unambiguously distinguishes among various 
theoretical predictions for the form factors of the $\gamma^*\gamma \ra \piz$ transition. 
We have demonstrated that the non-perturbative properties 
of $\piz$ and $\etaz$ agree with each other 
which indicates that the wave functions of these two mesons are similar. 
In the $\etap$ analysis we have shown that the non-perturbative properties 
of $\etap$ differ substantially from those of $\piz$ and $\etaz$. 
Our measurement for $\etap$ provides important information for 
future theoretical investigations of the structure of this particle. 

\section{Acknowledgements}

\smallskip
We gratefully acknowledge the effort of the CESR staff in providing us with
excellent luminosity and running conditions.
J.P.A., J.R.P., and I.P.J.S. thank                                           
the NYI program of the NSF, 
M.S. thanks the PFF program of the NSF,
G.E. thanks the Heisenberg Foundation, 
K.K.G., M.S., H.N.N., T.S., and H.Y. thank the
OJI program of DOE, 
J.R.P., K.H., M.S. and V.S. thank the A.P. Sloan Foundation,
A.W. and R.W. thank the 
Alexander von Humboldt Stiftung,
and M.S. thanks Research Corporation
for support.
This work was supported by the National Science Foundation, the
U.S. Department of Energy, and the Natural Sciences and Engineering Research 
Council of Canada.

\onecolumn

\begin{table}
\begin{center}
\caption{The results of the $\piz \ra \gaga$ analysis assuming ${\cal B} \equiv $ ${\cal B}(\piz \ra \gaga) = 0.99$. 
The differential cross section is for $e^+e^- \ra e^+e^-\piz$. 
}
\label{tab:piz_table_1}
\smallskip
\begin{tabular}{cccccccc}
$Q^2$ {\scriptsize interval}&
$N_{\rm \piz}$&
$N_{\rm \piz}$&
$\epsilon$&
${\cal B} \times N_{\piz}$&
$\tilde{Q}^2$&
$d \sigma/d Q^2 (\tilde{Q}^2)$&
$\tilde{Q}^2 |\fpizg(\tilde{Q}^2)|$ 
\\
($\gevt$)&
{\scriptsize detected}&
{\scriptsize signal}&
\% &
{\scriptsize produced}&
($\gevt$)&
(fb / $\gevt$) &
(0.01 $\times$ GeV) 
\\
\hline
 1.5 -- 1.8 & 150 $\pm$ 16 & 137 $\pm$ 17 & 7.5 & 1831 $\pm$ 231 & 1.64 & 2145 $\pm$ 270 $\pm$ 105 & 12.1 $\pm$ 0.8 $\pm$ 0.3  \\
 1.8 -- 2.0 & 174 $\pm$ 19 & 163 $\pm$ 20 & 24  &  686 $\pm$ 82  & 1.90 & 1205 $\pm$ 144 $\pm$  59 & 11.7 $\pm$ 0.7 $\pm$ 0.3  \\
 2.0 -- 2.2 & 193 $\pm$ 19 & 182 $\pm$ 20 & 26  &  688 $\pm$ 74  & 2.10 & 1209 $\pm$ 131 $\pm$  59 & 13.8 $\pm$ 0.8 $\pm$ 0.3  \\
 2.2 -- 2.4 & 125 $\pm$ 16 & 120 $\pm$ 16 & 28  &  424 $\pm$ 57  & 2.30 &  744 $\pm$ 100 $\pm$  37 & 12.7 $\pm$ 0.9 $\pm$ 0.3  \\
 2.4 -- 2.6 & 106 $\pm$ 15 & 101 $\pm$ 15 & 29  &  355 $\pm$ 52  & 2.50 &  624 $\pm$  92 $\pm$  31 & 13.5 $\pm$ 1.0 $\pm$ 0.3  \\
 2.6 -- 2.8 & 102 $\pm$ 14 & 99  $\pm$ 15 & 29  &  342 $\pm$ 50  & 2.70 &  602 $\pm$  89 $\pm$  30 & 15.1 $\pm$ 1.1 $\pm$ 0.4  \\
 2.8 -- 3.1 & 99  $\pm$ 15 & 88  $\pm$ 16 & 29  &  309 $\pm$ 56  & 2.94 &  362 $\pm$  65 $\pm$  18 & 13.7 $\pm$ 1.2 $\pm$ 0.3  \\
 3.1 -- 3.5 & 107 $\pm$ 15 & 97  $\pm$ 16 & 30  &  321 $\pm$ 53  & 3.29 &  282 $\pm$  47 $\pm$  14 & 14.5 $\pm$ 1.2 $\pm$ 0.4  \\
 3.5 -- 4.0 & 75  $\pm$ 13 & 65  $\pm$ 14 & 31  &  213 $\pm$ 46  & 3.74 &  150 $\pm$  32 $\pm$   7 & 13.2 $\pm$ 1.4 $\pm$ 0.3  \\
 4.0 -- 4.5 & 43  $\pm$ 10 & 43  $\pm$ 10 & 31  &  138 $\pm$ 31  & 4.24 &   97 $\pm$  22 $\pm$   5 & 13.4 $\pm$ 1.5 $\pm$ 0.3  \\
 4.5 -- 5.0 & 40  $\pm$ 9  & 40  $\pm$  9 & 33  &  122 $\pm$ 26  & 4.74 &   85 $\pm$  18 $\pm$   4 & 15.4 $\pm$ 1.7 $\pm$ 0.4  \\
 5.0 -- 5.5 & 26  $\pm$ 6  & 26  $\pm$  6 & 34  &   76 $\pm$ 18  & 5.24 &   54 $\pm$  13 $\pm$   3 & 14.5 $\pm$ 1.8 $\pm$ 0.4  \\
 5.5 -- 6.0 & 20  $\pm$ 6  & 20  $\pm$  6 & 32  &   63 $\pm$ 18  & 5.74 &   44 $\pm$  12 $\pm$   2 & 15.5 $\pm$ 2.2 $\pm$ 0.4  \\
 6.0 -- 7.0 & 23  $\pm$ 6  & 23  $\pm$  6 & 31  &   74 $\pm$ 20  & 6.47 &   26 $\pm$   7 $\pm$   1 & 14.8 $\pm$ 2.0 $\pm$ 0.4  \\
 7.0 -- 9.0 & 15  $\pm$ 5  & 15  $\pm$  5 & 16  &   94 $\pm$ 28  & 7.90 &   17 $\pm$   5 $\pm$   1 & 16.7 $\pm$ 2.5 $\pm$ 0.4  \\
\end{tabular}
\end{center}
\end{table}

\begin{table}
\begin{center}
\caption{The results of the $\etaz \ra \gaga$ analysis assuming ${\cal B} \equiv $ ${\cal B}(\etaz \ra \gaga) = 0.39$. 
The differential cross section is for $e^+e^- \ra e^+e^-\etaz$. 
}
\label{tab:etaz_table_1}
\smallskip
\begin{tabular}{cccccccc}
$Q^2$ {\scriptsize interval}&
$N_{\rm \etaz}$&
$N_{\rm \etaz}$&
$\epsilon$&
${\cal B} \times N_{\etaz}$&
$\tilde{Q}^2$&
$d \sigma/d Q^2 (\tilde{Q}^2)$&
$\tilde{Q}^2 |\fetazg(\tilde{Q}^2)|$ 
\\
($\gevt$)&
{\scriptsize detected}&
{\scriptsize signal}&
\% &
{\scriptsize produced}&
($\gevt$)&
(fb / $\gevt$) &
(0.01 $\times$ GeV) 
\\
\hline
 1.5 -- 2.0 & 73 $\pm$ 12 & 73 $\pm$ 12 &  9.4 & 768 $\pm$ 131 & 1.73 & 1359 $\pm$ 231 $\pm$ 67 & 10.9 $\pm$ 0.9 $\pm$ 0.3 \\
 2.0 -- 2.5 & 81 $\pm$ 14 & 81 $\pm$ 14 &  21  & 392 $\pm$  66 & 2.23 &  694 $\pm$ 117 $\pm$ 34 & 12.0 $\pm$ 1.0 $\pm$ 0.3 \\
 2.5 -- 3.0 & 59 $\pm$ 10 & 59 $\pm$ 10 &  22  & 264 $\pm$  47 & 2.74 &  467 $\pm$  83 $\pm$ 23 & 13.9 $\pm$ 1.2 $\pm$ 0.3 \\
 3.0 -- 3.5 & 35 $\pm$  8 & 35 $\pm$  8 &  25  & 142 $\pm$  33 & 3.24 &  251 $\pm$  59 $\pm$ 12 & 13.6 $\pm$ 1.6 $\pm$ 0.3 \\
 3.5 -- 4.0 & 19 $\pm$  7 & 19 $\pm$  7 &  24  &  78 $\pm$  29 & 3.74 &  138 $\pm$  51 $\pm$  7 & 12.8 $\pm$ 2.4 $\pm$ 0.3 \\
 4.0 -- 5.0 & 28 $\pm$  8 & 28 $\pm$  8 &  27  & 105 $\pm$  29 & 4.46 &   93 $\pm$  26 $\pm$  5 & 14.5 $\pm$ 2.0 $\pm$ 0.4 \\
 5.0 -- 6.5 & 22 $\pm$  6 & 22 $\pm$  6 &  28  &  79 $\pm$  22 & 5.68 &   47 $\pm$  13 $\pm$  2 & 15.7 $\pm$ 2.2 $\pm$ 0.4 \\
 6.5 -- 9.0 &  8 $\pm$  3 &  8 $\pm$  3 &  18  &  46 $\pm$  19 & 7.58 &   16 $\pm$   7 $\pm$  1 & 15.3 $\pm$ 3.2 $\pm$ 0.4 \\
\end{tabular}
\end{center}
\end{table}

\begin{table}
\begin{center}
\caption{The results of the $\etaz \ra 3\piz$ analysis assuming ${\cal B} \equiv $ ${\cal B}(\etaz \ra 3\piz) \times $ ${\cal B}^3(\piz \ra \gaga) = 0.31$. 
The differential cross section is for $e^+e^- \ra e^+e^-\etaz$. 
}
\label{tab:etaz_table_2}
\smallskip
\begin{tabular}{cccccccc}
$Q^2$ {\scriptsize interval}&
$N_{\rm \etaz}$&
$N_{\rm \etaz}$&
$\epsilon$&
${\cal B} \times N_{\etaz}$&
$\tilde{Q}^2$&
$d \sigma/d Q^2 (\tilde{Q}^2)$&
$\tilde{Q}^2 |\fetazg(\tilde{Q}^2)|$ 
\\
($\gevt$)&
{\scriptsize detected}&
{\scriptsize signal}&
\% &
{\scriptsize produced}&
($\gevt$)&
(fb / $\gevt$) &
(0.01 $\times$ GeV) 
\\
\hline
 1.5 -- 2.0 & 39 $\pm$ 7 & 37 $\pm$ 7 & 6.9 & 544 $\pm$ 95 & 1.73 & 1219 $\pm$ 212 $\pm$  90 & 10.3 $\pm$ 0.9 $\pm$ 0.4 \\
 2.0 -- 2.5 & 57 $\pm$ 8 & 54 $\pm$ 8 & 14  & 392 $\pm$ 57 & 2.23 &  879 $\pm$ 128 $\pm$  65 & 13.5 $\pm$ 1.0 $\pm$ 0.5 \\
 2.5 -- 3.5 & 47 $\pm$ 7 & 45 $\pm$ 7 & 16  & 279 $\pm$ 44 & 2.94 &  312 $\pm$  50 $\pm$  23 & 12.9 $\pm$ 1.0 $\pm$ 0.4 \\
 3.5 -- 5.6 & 24 $\pm$ 5 & 24 $\pm$ 5 & 18  & 132 $\pm$ 31 & 4.16 &   99 $\pm$  23 $\pm$   7 & 13.1 $\pm$ 1.5 $\pm$ 0.4 \\
 5.6 -- 9.0 & 20 $\pm$ 5 & 20 $\pm$ 5 & 15  & 135 $\pm$ 34 & 6.56 &   38 $\pm$  10 $\pm$   3 & 18.3 $\pm$ 2.3 $\pm$ 0.7 \\
\end{tabular}
\end{center}
\end{table}

\begin{table}
\begin{center}
\caption{The results of the $\etaz \ra \pip\pim\piz$ analysis assuming ${\cal B} \equiv $ ${\cal B}(\etaz \ra \pip\pim\piz) \times $ ${\cal B}(\piz \ra \gaga) = 0.23$.
The differential cross section is for $e^+e^- \ra e^+e^-\etaz$. 
}
\label{tab:etaz_table_3}
\smallskip
\begin{tabular}{cccccccc}
$Q^2$ {\scriptsize interval}&
$N_{\rm \etaz}$&
$N_{\rm \etaz}$&
$\epsilon$&
${\cal B} \times N_{\etaz}$&
$\tilde{Q}^2$&
$d \sigma/d Q^2 (\tilde{Q}^2)$&
$\tilde{Q}^2 |\fetazg(\tilde{Q}^2)|$ 
\\
($\gevt$)&
{\scriptsize detected}&
{\scriptsize signal}&
\% &
{\scriptsize produced}&
($\gevt$)&
(fb / $\gevt$) &
(0.01 $\times$ GeV) 
\\
\hline
 1.5 --  2.0 & 37 $\pm$ 6 & 37 $\pm$ 6 & 10 & 385 $\pm$ 67 &  1.73 & 1167 $\pm$ 202 $\pm$  90 & 10.1 $\pm$ 0.87 $\pm$ 0.39 \\
 2.0 --  2.5 & 51 $\pm$ 7 & 50 $\pm$ 7 & 21 & 235 $\pm$ 35 &  2.23 &  714 $\pm$ 105 $\pm$  55 & 12.1 $\pm$ 0.89 $\pm$ 0.47 \\
 2.5 --  3.5 & 49 $\pm$ 7 & 48 $\pm$ 7 & 23 & 210 $\pm$ 31 &  2.94 &  318 $\pm$  47 $\pm$  25 & 13.0 $\pm$ 0.96 $\pm$ 0.50 \\
 3.5 --  5.0 & 31 $\pm$ 6 & 31 $\pm$ 6 & 26 & 117 $\pm$ 23 &  4.16 &  118 $\pm$  23 $\pm$   9 & 14.4 $\pm$ 1.39 $\pm$ 0.55 \\
 5.0 --  9.0 & 32 $\pm$ 6 & 32 $\pm$ 6 & 26 & 122 $\pm$ 23 &  6.56 &   46 $\pm$   9 $\pm$   4 & 20.1 $\pm$ 1.88 $\pm$ 0.77 \\
 9.0 -- 20.0 &  6 $\pm$ 3 &  6 $\pm$ 3 & 25 &  23 $\pm$ 10 & 12.74 &  3.1 $\pm$ 1.4 $\pm$ 0.2 & 18.4 $\pm$ 4.19 $\pm$ 0.71 \\
\end{tabular}
\end{center}
\end{table}

\begin{table}
\begin{center}
\caption{The results of the $\etap \ra \piz\piz\etaz \ra 6\gamma$ analysis assuming 
${\cal B} \equiv $ ${\cal B}(\etap \ra \piz\piz\etaz) \times {\cal B}(\etaz \ra \gaga) \times $ ${\cal B}^2(\piz \ra \gaga) = 0.080$. 
The differential cross section is for $e^+e^- \ra e^+e^-\etap$. 
}
\label{tab:etap_table_1}
\smallskip
\begin{tabular}{cccccccc}
$Q^2$ {\scriptsize interval}&
$N_{\rm \etap}$&
$N_{\rm \etap}$&
$\epsilon$&
${\cal B} \times N_{\etap}$&
$\tilde{Q}^2$&
$d \sigma/d Q^2 (\tilde{Q}^2)$&
$\tilde{Q}^2 |\fetapg(\tilde{Q}^2)|$ 
\\
($\gevt$)&
{\scriptsize detected}&
{\scriptsize signal}&
\% &
{\scriptsize produced}&
($\gevt$)&
(fb / $\gevt$) &
(0.01 $\times$ GeV) 
\\
\hline
 1.5 -- 2.0 & 40 $\pm$ 7 & 40 $\pm$ 7 & 8.4 & 474 $\pm$ 85 & 1.73 & 4132 $\pm$ 740 $\pm$ 310 & 20.1 $\pm$ 1.8 $\pm$ 0.8 \\
 2.0 -- 2.5 & 40 $\pm$ 7 & 40 $\pm$ 7 & 16  & 259 $\pm$ 44 & 2.23 & 2258 $\pm$ 381 $\pm$ 169 & 22.7 $\pm$ 1.9 $\pm$ 0.9 \\
 2.5 -- 3.5 & 29 $\pm$ 6 & 29 $\pm$ 6 & 16  & 176 $\pm$ 38 & 2.94 &  767 $\pm$ 164 $\pm$  58 & 21.1 $\pm$ 2.3 $\pm$ 0.8 \\
 3.5 -- 5.0 & 17 $\pm$ 4 & 17 $\pm$ 4 & 18  &  94 $\pm$ 24 & 4.16 &  274 $\pm$  70 $\pm$  21 & 22.7 $\pm$ 2.9 $\pm$ 0.9 \\
 5.0 -- 9.0 & 14 $\pm$ 4 & 14 $\pm$ 4 & 16  &  90 $\pm$ 24 & 6.56 &   98 $\pm$  26 $\pm$   7 & 30.0 $\pm$ 4.0 $\pm$ 1.1 \\
\end{tabular}
\end{center}
\end{table}

\begin{table}
\begin{center}
\caption{The results of the $\etap \ra \piz\piz\etaz \ra 5\piz \ra 10\gamma$ analysis assuming 
${\cal B} \equiv $ ${\cal B}(\etap \ra \piz\piz\etaz) \times $ ${\cal B}(\etaz \ra 3\piz) \times $ ${\cal B}^5(\piz \ra \gaga) = 0.063$. 
The differential cross section is for $e^+e^- \ra e^+e^-\etap$. 
}
\label{tab:etap_table_2}
\smallskip
\begin{tabular}{cccccccc}
$Q^2$ {\scriptsize interval}&
$N_{\rm \etap}$&
$N_{\rm \etap}$&
$\epsilon$&
${\cal B} \times N_{\etap}$&
$\tilde{Q}^2$&
$d \sigma/d Q^2(\tilde{Q}^2)$&
$\tilde{Q}^2 |\fetapg(\tilde{Q}^2)|$ 
\\
($\gevt$)&
{\scriptsize detected}&
{\scriptsize signal}&
\% &
{\scriptsize produced}&
($\gevt$)&
(fb / $\gevt$) &
(0.01 $\times$ GeV) 
\\
\hline
 1.5 -- 3.0 & 18 $\pm$ 5 & 18 $\pm$ 5 & 3.5 & 510 $\pm$ 153 & 2.09 & 1875 $\pm$ 563 $\pm$ 204 & 18.6 $\pm$ 2.8 $\pm$ 1.0 \\
 3.0 -- 9.0 &  7 $\pm$ 3 &  7 $\pm$ 2 & 5.4 & 129 $\pm$  49 & 4.92 &  118 $\pm$  45 $\pm$  13 & 20.0 $\pm$ 3.8 $\pm$ 1.1 \\
\end{tabular}
\end{center}
\end{table}

\begin{table}
\begin{center}
\caption{The results of the $\etap \ra \rhoz\gamma \ra \pip\pim\gamma$ analysis assuming 
${\cal B} \equiv $ ${\cal B}(\etap \ra \rhoz\gamma) \times $ ${\cal B}(\rhoz \ra \pip\pim) = 0.30$. 
The differential cross section is for $e^+e^- \ra e^+e^-\etap$. 
}
\label{tab:etap_table_3}
\smallskip
\begin{tabular}{cccccccc}
$Q^2$ {\scriptsize interval}&
$N_{\rm \etap}$&
$N_{\rm \etap}$&
$\epsilon$&
${\cal B} \times N_{\etap}$&
$\tilde{Q}^2$&
$d \sigma/d Q^2 (\tilde{Q}^2)$&
$\tilde{Q}^2 |\fetapg(\tilde{Q}^2)|$ 
\\
($\gevt$)&
{\scriptsize detected}&
{\scriptsize signal}&
\% &
{\scriptsize produced}&
($\gevt$)&
(fb / $\gevt$) &
(0.01 $\times$ GeV) 
\\
\hline
 1.5 --  2.0 & 111 $\pm$ 13 & 111 $\pm$ 13 & 8.9 & 1257 $\pm$ 152 &  1.73 & 2891 $\pm$ 350 $\pm$ 197 & 16.8 $\pm$ 1.02 $\pm$ 0.57 \\
 2.0 --  2.5 & 131 $\pm$ 14 & 131 $\pm$ 14 & 17  &  765 $\pm$  84 &  2.23 & 1759 $\pm$ 193 $\pm$ 120 & 20.0 $\pm$ 1.10 $\pm$ 0.68 \\
 2.5 --  3.5 & 123 $\pm$ 14 & 123 $\pm$ 14 & 21  &  593 $\pm$  69 &  2.94 &  681 $\pm$  79 $\pm$  46 & 19.9 $\pm$ 1.15 $\pm$ 0.68 \\
 3.5 --  5.0 &  86 $\pm$ 11 &  86 $\pm$ 11 & 24  &  353 $\pm$  47 &  4.16 &  270 $\pm$  36 $\pm$  18 & 22.6 $\pm$ 1.51 $\pm$ 0.77 \\
 5.0 --  9.0 &  49 $\pm$ 10 &  49 $\pm$ 10 & 31  &  158 $\pm$  32 &  6.56 &   45 $\pm$   9 $\pm$   3 & 20.4 $\pm$ 2.08 $\pm$ 0.69 \\
 9.0 -- 30.0 &  22 $\pm$  8 &  22 $\pm$  8 & 37  &   58 $\pm$  21 & 15.30 &  3.2 $\pm$ 1.1 $\pm$ 0.2 & 24.8 $\pm$ 4.44 $\pm$ 0.84 \\
\end{tabular}
\end{center}
\end{table}

\begin{table}
\begin{center}
\caption{The results of the $\etap \ra \pip\pim\etaz \ra \pip\pim2\gamma$ analysis assuming 
${\cal B} \equiv $ ${\cal B}(\etap \ra \etaz\pip\pim) \times $ ${\cal B}(\etaz \ra \gaga) = 0.17$. 
The differential cross section is for $e^+e^- \ra e^+e^-\etap$. 
}
\label{tab:etap_table_4}
\smallskip
\begin{tabular}{cccccccc}
$Q^2$ {\scriptsize interval}&
$N_{\rm \etap}$&
$N_{\rm \etap}$&
$\epsilon$&
${\cal B} \times N_{\etap}$&
$\tilde{Q}^2$&
$d \sigma/d Q^2 (\tilde{Q}^2)$&
$\tilde{Q}^2 |\fetapg(\tilde{Q}^2)|$ 
\\
($\gevt$)&
{\scriptsize detected}&
{\scriptsize signal}&
\% &
{\scriptsize produced}&
($\gevt$)&
(fb / $\gevt$) &
(0.01 $\times$ GeV) 
\\
\hline
 1.5 --  2.0 & 57 $\pm$ 8 & 57 $\pm$ 8 & 7.6 & 743 $\pm$ 104 &  1.73 & 3007 $\pm$ 421  $\pm$ 199 & 17.1 $\pm$ 1.2 $\pm$ 0.6 \\
 2.0 --  2.5 & 70 $\pm$ 9 & 70 $\pm$ 9 & 17  & 408 $\pm$  51 &  2.23 & 1651 $\pm$ 208  $\pm$ 109 & 19.4 $\pm$ 1.2 $\pm$ 0.6 \\
 2.5 --  3.5 & 60 $\pm$ 8 & 60 $\pm$ 8 & 21  & 282 $\pm$  38 &  2.94 &  570 $\pm$  77  $\pm$  38 & 18.2 $\pm$ 1.2 $\pm$ 0.6 \\
 3.5 --  5.0 & 58 $\pm$ 8 & 58 $\pm$ 8 & 27  & 216 $\pm$  30 &  4.16 &  292 $\pm$  40  $\pm$  19 & 23.5 $\pm$ 1.6 $\pm$ 0.8 \\
 5.0 --  9.0 & 45 $\pm$ 7 & 45 $\pm$ 7 & 34  & 133 $\pm$  20 &  6.56 &   67 $\pm$  10  $\pm$   4 & 24.9 $\pm$ 1.9 $\pm$ 0.8 \\
 9.0 -- 30.0 & 16 $\pm$ 4 & 16 $\pm$ 4 & 36  &  44 $\pm$  11 & 15.30 &  4.3 $\pm$ 1.1  $\pm$ 0.3 & 28.8 $\pm$ 3.6 $\pm$ 0.9 \\
\end{tabular}
\end{center}
\end{table}

\begin{table}
\begin{center}
\caption{The results of the $\etap \ra \pip\pim\etaz \ra 2\pip2\pim\piz \ra 2\pip2\pim2\gamma$ analysis assuming 
${\cal B} \equiv $ ${\cal B}(\etap \ra \pip\pim\etaz) \times $ ${\cal B}(\etaz \ra \pip\pim\piz) \times $ ${\cal B}(\piz \ra \gaga) = 0.10$. 
The differential cross section is for $e^+e^- \ra e^+e^-\etap$. 
}
\label{tab:etap_table_5}
\smallskip
\begin{tabular}{cccccccc}
$Q^2$ {\scriptsize interval}&
$N_{\rm \etap}$&
$N_{\rm \etap}$&
$\epsilon$&
${\cal B} \times N_{\etap}$&
$\tilde{Q}^2$&
$d \sigma/d Q^2 (\tilde{Q}^2)$&
$\tilde{Q}^2 |\fetapg(\tilde{Q}^2)|$ 
\\
($\gevt$)&
{\scriptsize detected}&
{\scriptsize signal}&
\% &
{\scriptsize produced}&
($\gevt$)&
(fb / $\gevt$) &
(0.01 $\times$ GeV) 
\\
\hline
 1.5 --  2.5 & 33 $\pm$ 6 & 33 $\pm$ 6 & 6.2 & 528 $\pm$ 95 &  1.92 & 1830 $\pm$ 329 $\pm$ 176 & 15.9 $\pm$ 1.4 $\pm$ 0.7 \\
 2.5 --  3.5 & 22 $\pm$ 5 & 22 $\pm$ 5 & 13  & 169 $\pm$ 40 &  2.94 &  584 $\pm$ 138 $\pm$  56 & 18.4 $\pm$ 2.2 $\pm$ 0.9 \\
 3.5 --  5.0 & 18 $\pm$ 5 & 18 $\pm$ 5 & 16  & 113 $\pm$ 30 &  4.16 &  261 $\pm$  69 $\pm$  25 & 22.2 $\pm$ 2.9 $\pm$ 1.1 \\
 5.0 --  9.0 & 15 $\pm$ 4 & 15 $\pm$ 4 & 21  &  74 $\pm$ 20 &  6.56 &   64 $\pm$  17 $\pm$   6 & 24.4 $\pm$ 3.2 $\pm$ 1.2 \\
 9.0 -- 30.0 &  4 $\pm$ 2 &  4 $\pm$ 2 & 24  &  16 $\pm$  8 & 15.30 &  2.7 $\pm$ 1.4 $\pm$ 0.3 & 22.9 $\pm$ 5.8 $\pm$ 1.1 \\
\end{tabular}
\end{center}
\end{table}

\begin{table}
\begin{center}
\caption{The results of the $\etap \ra \pip\pim3\piz \ra \pip\pim6\gamma$ analysis assuming ${\cal B} \equiv 
$ $( 
{\cal B}(\etap \ra \pip\pim\etaz) \times $ ${\cal B}(\etaz \ra 3\piz) + 
$ ${\cal B}(\etap \ra \piz\piz\etaz) \times $ ${\cal B}(\etaz \ra \pip\pim\piz)
) 
\times {\cal B}^3(\piz \ra \gaga) = 0.14$. 
The differential cross section is for $e^+e^- \ra e^+e^-\etap$. 
}
\label{tab:etap_table_6}
\smallskip
\begin{tabular}{cccccccc}
$Q^2$ {\scriptsize interval}&
$N_{\rm \etap}$&
$N_{\rm \etap}$&
$\epsilon$&
${\cal B} \times N_{\etap}$&
$\tilde{Q}^2$&
$d \sigma/d Q^2 (\tilde{Q}^2)$&
$\tilde{Q}^2 |\fetapg(\tilde{Q}^2)|$ 
\\
($\gevt$)&
{\scriptsize detected}&
{\scriptsize signal}&
\% &
{\scriptsize produced}&
($\gevt$)&
(fb / $\gevt$) &
(0.01 $\times$ GeV) 
\\
\hline
 1.5 --  2.5 & 54 $\pm$ 8 & 54 $\pm$ 8 & 3.7 & 1468 $\pm$ 206 &  1.92 & 2803 $\pm$ 393 $\pm$ 247 & 19.7 $\pm$ 1.4 $\pm$ 0.9 \\
 2.5 --  3.5 & 25 $\pm$ 6 & 25 $\pm$ 6 & 7.5 &  330 $\pm$  78 &  2.94 &  630 $\pm$ 149 $\pm$  55 & 19.1 $\pm$ 2.3 $\pm$ 0.8 \\
 3.5 --  5.0 & 15 $\pm$ 4 & 15 $\pm$ 4 &  10 &  161 $\pm$  45 &  4.16 &  205 $\pm$  57 $\pm$  18 & 19.7 $\pm$ 2.7 $\pm$ 0.9 \\
 5.0 --  9.0 & 13 $\pm$ 4 & 13 $\pm$ 4 &  13 &  101 $\pm$  34 &  6.56 &   48 $\pm$  16 $\pm$   4 & 21.0 $\pm$ 3.5 $\pm$ 0.9 \\
 9.0 -- 30.0 &  2 $\pm$ 1 &  2 $\pm$ 1 &  15 &   14 $\pm$  10 & 15.30 &  1.3 $\pm$ 0.9 $\pm$ 0.1 & 15.7 $\pm$ 5.6 $\pm$ 0.7 \\
\end{tabular}
\end{center}
\end{table}

\begin{table}
\begin{center}
\caption{Values of the pole-mass parameters 
$\Lambda_{\piz}$, $\Lambda_{\etaz}$ and $\Lambda_{\etap}$ 
measured using various final states. 
For each measurement, the first error is statistical, 
the second error represents the systematic uncertainties of our measurement 
and the third error reflects the experimental 
error in the value of the two-photon partial width of the meson. 
}
\label{tab:final_res_pole}
\smallskip
\begin{tabular}{cc}
Decay chain & $\Lambda_{\cal R}$ (MeV) \\
\hline
$\piz\ra \gaga$ 				& 776 $\pm$ 10 $\pm$ 12 $\pm$ 16 \\
\hline
\hline
$\etaz \ra \gaga$ 				& 778 $\pm$ 19 $\pm$ 12 $\pm$ 22 \\
$\etaz \ra 3\piz \ra 6\gamma$			& 773 $\pm$ 20 $\pm$ 17 $\pm$ 22 \\
$\etaz \ra \pip\pim\piz \ra \pip\pim2\gamma$ 	& 773 $\pm$ 18 $\pm$ 18 $\pm$ 22 \\
\hline
Simultaneous fit to all $\etaz$ data 		& 774 $\pm$ 11 $\pm$ 16 $\pm$ 22 \\
\hline
\hline
$\etap \ra \rhoz\gamma \ra \pip\pim\gamma$ 	& 857 $\pm$ 15 $\pm$ 19 $\pm$ 19 \\
$\etap \ra \pip\pim\etaz \ra \pip\pim2\gamma$ 	& 864 $\pm$ 16 $\pm$ 18 $\pm$ 19 \\
$\etap \ra \pip\pim3\piz \ra \pip\pim6\gamma$ 	& 838 $\pm$ 27 $\pm$ 21 $\pm$ 17 \\
$\etap \ra \pip\pim\etaz \ra 2\pip2\pim2\gamma$ & 824 $\pm$ 29 $\pm$ 25 $\pm$ 18 \\
$\etap \ra \piz\piz\etaz \ra 6\gamma$ 		& 931 $\pm$ 29 $\pm$ 21 $\pm$ 23 \\
$\etap \ra \piz\piz\etaz \ra 10\gamma$ 		& 837 $\pm$ 61 $\pm$ 27 $\pm$ 17 \\
\hline
Simultaneous fit to all $\etap$ data 		& 859 $\pm$ 9 $\pm$ 18 $\pm$ 20 \\
\end{tabular}
\end{center}
\end{table}

% ------------------------------------ FIGURES ------------------------------------

\twocolumn

%\newpage

%===================  MASS PLOTS GO HERE ===================

%===================  MASS PLOTS DO NOT GO HERE ===================

%===================  MASS PLOTS GO HERE ===================

%+++++++++++++++++++++++++++++++++++++++++++++++++++++++++++++

%----------------- CROSS SECTIONS -----------------

%------------- RESULTS RESULTS RESULTS RESULTS -------------

\end{document}